\documentclass[pdflatex,sn-mathphys-num]{sn-jnl}
\usepackage[left]{lineno}

\usepackage{amsmath,amssymb,amsfonts}%
\usepackage{amsthm}%
\usepackage[title]{appendix}%
\usepackage{xcolor}%
\usepackage{textcomp}%
\usepackage{manyfoot}%
\usepackage{booktabs}%
\usepackage{algorithm}%
\usepackage{algorithmicx}%
\usepackage{algpseudocode}%
\usepackage{listings}%

\usepackage[section]{placeins}
\usepackage{tabularx}
\usepackage{multirow}
\usepackage{CJKutf8}
\graphicspath{{figures/}} 

\theoremstyle{thmstyleone}%
\theoremstyle{thmstyletwo}%

\theoremstyle{thmstylethree}%

\usepackage{derivative}

\newcommand*{\addFileDependency}[1]{
\typeout{(#1)}
\@addtofilelist{#1}
\IfFileExists{#1}{}{\typeout{No file #1.}}
}\makeatother

\begin{document}
\title[FuXi-TC]{FuXi-TC: A generative framework integrating deep learning and physics-based models for improved tropical cyclone forecasts}

\author[2]{\fnm{Shan} \sur{Guo}}\email{guoshan@sais.org.cn}
\equalcont{These authors contributed equally to this work.}

\author[1,2]{\fnm{Lei} \sur{Chen}}\email{cltpys@163.com}
\equalcont{These authors contributed equally to this work.}

\author[1,2]{\fnm{Yangyang} \sur{Zhao}}\email{zhaoyangyang@sais.org.cn}

\author[1]{\fnm{Yuetan} \sur{Lin}}\email{linyuetan@sais.org.cn}

\author[3,4]{\fnm{Zeyi} \sur{Niu}}\email{niuzy@typhoon.org.cn}

\author[3]{\fnm{Xinyan} \sur{Zhang}}\email{zhangxy@typhoon.org.cn}

\author[3,4]{\fnm{Ziyao} \sur{Sun}}\email{sunzy@typhoon.org.cn}

\author*[1,2]{\fnm{Xiaohui} \sur{Zhong}}\email{x7zhong@gmail.com}

\author*[1,2]{\fnm{Hao} \sur{Li}}\email{lihao$\_$lh@fudan.edu.cn}

\affil[1]{\orgdiv{Artificial Intelligence Innovation and Incubation Institute}, \orgname{Fudan University}, \orgaddress{\city{Shanghai}, \postcode{200433}, \country{China}}}

\affil[2]{\orgname{Shanghai Academy of Artificial Intelligence for Science}, \orgaddress{\city{Shanghai}, \postcode{200232}, \country{China}}}

\affil[3]{\orgdiv{Department of Atmospheric and Oceanic Sciences, Institute of Atmospheric Sciences}, \orgname{Fudan University}, \orgaddress{\city{Shanghai}, \postcode{200433}, \country{China}}}

\affil[4]{\orgdiv{Shanghai Typhoon Institute}, \orgname{China Meteorological Administration}, \orgaddress{\city{Shanghai}, \postcode{200030}, \country{China}}}


\abstract{
Tropical cyclones (TCs) are among the most devastating natural hazards, yet their intensity remains notoriously difficult to predict.
Numerical weather prediction (NWP) models are constrained by both computational demands and intrinsic predictability, while state-of-the-art deep learning-based weather forecasting models tend to underestimate TC intensity due to biases in reanalysis-based training data.Here, we present FuXi-TC, a diffusion-based generative forecasting framework that combines the track prediction strength of the FuXi model with the intensity representation of NWP simulations. By conditioning a diffusion model on the large-scale forecasts of the global FuXi model, FuXi-TC effectively downscales and delivers higher-accuracy forecasts of fine-grained variable fields such as wind speed and precipitation. In evaluations across the 2024 Western North Pacific, our approach matches the TC intensity forecast skill of the operational  European Centre for Medium-Range Weather Forecasts (ECMWF) deterministic model while delivering superior precipitation forecasts. Meanwhile this is achieved with significantly higher inference speeds and lower computational costs.
Moreover, FuXi-TC demonstrates robust zero-shot generalization directly when applied to North Atlantic hurricanes without any fine-tuning or re-training. When applied to the FuXi ensemble model, this framework effectively yields well-dispersed probabilistic forecasts and refines the ensemble intensity predictions. More broadly, these results highlight the potential of leveraging physics-based simulations to construct new training datasets that further advancing deep learning–based weather forecasting.
}


\maketitle

\section{Introduction}
Tropical cyclones (TCs) frequently inflict catastrophic damage on coastal and inland regions through severe winds, heavy rainfall, storm surges, and widespread flooding.
They pose serious threats to infrastructure, economic activities, and human safety across affected areas ~\cite{emanuel2005increasing,emanuel2012impact,zhang2009tropical,zhang2017impact,peduzzi2012global}.
Under global warming, although the global annual frequency of TCs has remained relatively stable, the occurrence and spatial extent of extreme-intensity events have increased \cite{IPCC_AR6_WGI_2021}.
Continued ocean warming is expected to further intensify TC strength, enhance precipitation, and prolong cyclone lifespans, amplifying societal risk and increasing demands on disaster prevention and mitigation systems \cite{webster2005changes}.

Accurate and timely prediction of TC track and intensity remains a critical challenge for modern meteorological services.
Operational forecasts primarily rely on traditional numerical weather prediction (NWP) models specifically optimized for tropical cyclones. These include the Hurricane Weather Research and Forecasting (HWRF) model \cite{tallapragada2016hwrf,alaka2024hwrf} from the National Oceanic and Atmospheric Administration (NOAA), Hurricane Analysis and Forecast System (HAFS) \cite{wang2023hafs} from the National Centers for Environmental Prediction (NCEP), and China Meteorological Administration Typhoon Model (CMA-TYM) \cite{ma2021grapes}.
These models solve atmospheric dynamical and thermodynamical equations to simulate the evolution of large-scale circulation and weather systems from prescribed initial and boundary conditions.
However, their forecast accuracy is constrained by uncertainties in initial conditions, errors in physical parameterizations, and limited spatial resolution. 
Furthermore, high-resolution NWP simulations are computationally expensive, typically requiring several hours to complete, which limits their utility for rapid-response forecasting during extreme weather events.
Recent studies suggest that the predictability of TC track and intensity based on traditional NWP models may be approaching theoretical limits \cite{cangialosi2020progress,xu2025exploring}, with marginal returns from further increase in resolution.

In recent years, data-driven weather forecasting models have demonstrated strong skill in TC track prediction \cite{xinyuan2025review}.
For instance, Pangu-Weather \cite{bi2023accurate} outperformed the high-resolution forecasts (HRES) of the European Centre for Medium-Range Weather Forecasts (ECMWF) in track prediction accuracy for 88 TCs in 2018 .
Similarly, FuXi-Extreme \cite{zhong2024fuxi-exteme}, evaluated across 90 TCs, achieved superior performance in both TC track prediction and large-scale circulation patterns in influencing TC motion, particularly at lead times beyond 48 hours.
These systems offer fast inference and scalability, making them attractive for real-time applications \cite{mahesh2023evaluating,wang2024machine}.
Despite these advances, TC intensity prediction remains a key limitation of current data-driven models \cite{xu2025exploring}.
Artificial intelligence (AI) models are typically optimized for global statistical performance and tend to smooth extremes, leading to systematic underestimation of maximum winds \cite{shi2025comparison,zhong2024fuxi-exteme,liu2024hybrid}.
This issue is compounded by dependence on reanalysis datasets such as ERA5 for training \cite{hersbach2020era5}, which themselves underrepresent TC intensity due to resolution constraints and assimilation processes \cite{Hodges2017,zhang2024typhoon}.
Efforts to improve intensity forecasts using auxiliary datasets, such as International Best Track Archive for Climate Stewardship (IBTrACS) \cite{knapp2010international}, have primarily focused on predicting scalar attributes such as maximum sustained wind speed or minimum central pressure.
While valuable, these approaches do not directly enhance the full spatiotemporal structure of the TCs, which is critical for hazards such as wind damage and rainfall distribution.
Wang et al. \cite{wang2025vqlti} improved long-term tropical cyclone intensity forecasting by jointly leveraging IBTrACS intensity records and ERA5 reanalysis data. 
The model learns a discrete representation of historical intensity conditioned on surrounding ERA5 atmospheric fields, and performs temporal prediction in this latent space. 
Forecasts from the FengWu model are incorporated as conditioning information, and potential intensity calculated from the forecast fields is introduced to mitigate systematic underestimation and reduce long-term error accumulation.
Similarly, Huang et al. \cite{huang2025benchmark} constructed a global multimodal dataset by aligning best-track intensity records with meteorological reanalysis fields and environmental variables. By explicitly modeling environmental contexts such as the subtropical high, this method significantly improves track and intensity predictions.
Although both frameworks ingest complex multi-dimensional spatial meteorological grids as inputs, their final predictions are limited to time-series sequences of TC intensity. 
Consequently, these models fail to forecast the full spatiotemporal structure of the TC, such as the asymmetric wind fields or precipitation distributions, which is critical for accurately assessing localized hazards like wind damage and storm surges.

Hybrid strategies have attempted to integrate AI model forecasts with physical models to improve TC intensity prediction. These approaches leverage the superior large-scale predictive skill of AI models to provide improved initial and boundary conditions, while relying on physics-based NWP models to simulate fine-scale intensity evolution and spatiotemporal structures.
Xu et al.\cite{xu2025exploring} proposed a novel framework that uses the Pangu-Weather forecasts to drive the high-resolution Weather Research and Forecasting (WRF) model, augmented with vortex dynamic initialization. This approach improves typhoon intensity forecasts and demonstrates the promising potential of AI models to serve as superior large-scale forcings for physics-based regional forecasting.
Liu et al. \cite{liu2024hybrid} developed a novel hybrid forecasting framework that downscales Pangu-Weather forecasts with a high-resolution WRF model, incorporating large-scale spectral nudging and ocean mixed-layer coupling to extend tropical cyclone prediction to two weeks. This framework advances extended-range TC intensity and structural forecasting capability compared with AI models.
Syed et al. \cite{husain2025leveraging} proposes a hybrid NWP-AI forecasting system that employs large-scale spectral nudging to constrain the physics-based Global Environmental Multiscale model toward the superior large-scale predictions of the GraphCast model. This approach successfully leverages the strengths of both paradigms, and resulting in improved forecast skill across variables and lead times.
However, such hybrid approaches still depend on regional high-resolution NWP model for dynamical downscaling.
The high computational cost and slow integration speed of traditional NWP remain major bottlenecks for operational deployment.
A method that improves TC intensity and structure while preserving the computational efficiency of pure data-driven models is still lacking.

In this study, we present a fully data-driven framework for the rapid refinement of TC structure and rainfall forecasts, while preserving the inherent advantages of large meteorological models in accurate TC track prediction.
Specifically, we introduce FuXi-TC, a lightweight downscaling and refinement module based on a denoising diffusion probabilistic model (DDPM) \cite{sohl2015deep,ho2020denoising} that enhances 5-day forecasts produced by FuXi \cite{chen2023fuxi}.
Unlike the aforementioned hybrid approaches that require computationally expensive NWP models during the inference phase, our framework decouples the physical simulation from the real-time forecasting process. 
To address the intensity underestimation inherent in ERA5 and the limitation of IBTrACS-based methods that optimize scalar sequences, a key component of our approach is a physics-guided data construction strategy.
By generating multi-variable meteorological fields using WRF simulations as training targets, this strategy enables the trained model to reconstruct and refine the entire spatiotemporal structure of the TC.
Crucially, these high-resolution target fields are generated using WRF simulations dynamically downscaled from FuXi forecasts rather than from ERA5.
This design preserves structural and spatial consistency between the input FuXi forecasts and target fields, minimizing spatial misalignment arising from track deviations.
The resulting paired dataset facilitates the diffusion-based FuXi-TC model to focus on structural refinement and downscaling, rather than compensating for positional discrepancies.
The proposed framework is lightweight and highly adaptable, allowing for robust cross-basin generalization and extension to other existing large meteorological models such as FuXi-ENS\cite{zhong2024fuxiens}.
By combining physics-guided target generation with generative refinement, our approach advances TC intensity prediction while retaining the speed and scalability of AI models, offering a practical solution for operational deployment.

\section{Results}
For evaluating the comprehensive performance of the proposed FuXi-TC framework, our assessment encompasses several key dimensions.
First, we analyze the overall deterministic forecast performance in the Western North Pacific (WNP), focusing on TC track, intensity, and precipitation metrics, with supplementary evaluations on mean sea-level pressure (MSL; Supplementary Fig. 8) and additional precipitation statistics (Supplementary Figs. 5 and 6) provided in the supplementary materials.
Second, we present in-depth TC case studies of extreme events to illustrate the model's capability in capturing fine-scale wind structures and rainfall distribution.
Third, we evaluate the probabilistic forecasting performance by integrating FuXi‑TC into the FuXi‑ENS, resulting in an enhanced ensemble framework termed FuXi-TC-ENS. We analyze its ensemble spread characteristics and intensity estimation skill to assess uncertainty representation and probabilistic reliability. 
Fourth, we demonstrate the model's zero-shot geographical generalization by extending the evaluation to the North Atlantic (NA) basin, supported by specific hurricane case studies and domain configurations detailed in the supplementary materials.
Finally, we assess the computational efficiency of our approach relative to traditional NWP models.
Furthermore, additional evaluations, such as physical consistency using kinetic energy spectra to verify the effective resolution and spatial variance of the model outputs, are provided in supplementary information (Supplementary Fig. 7).

\subsection{Overall TC forecast performance}
\begin{figure}[htbp]
    \centering
    \includegraphics[width=\linewidth]{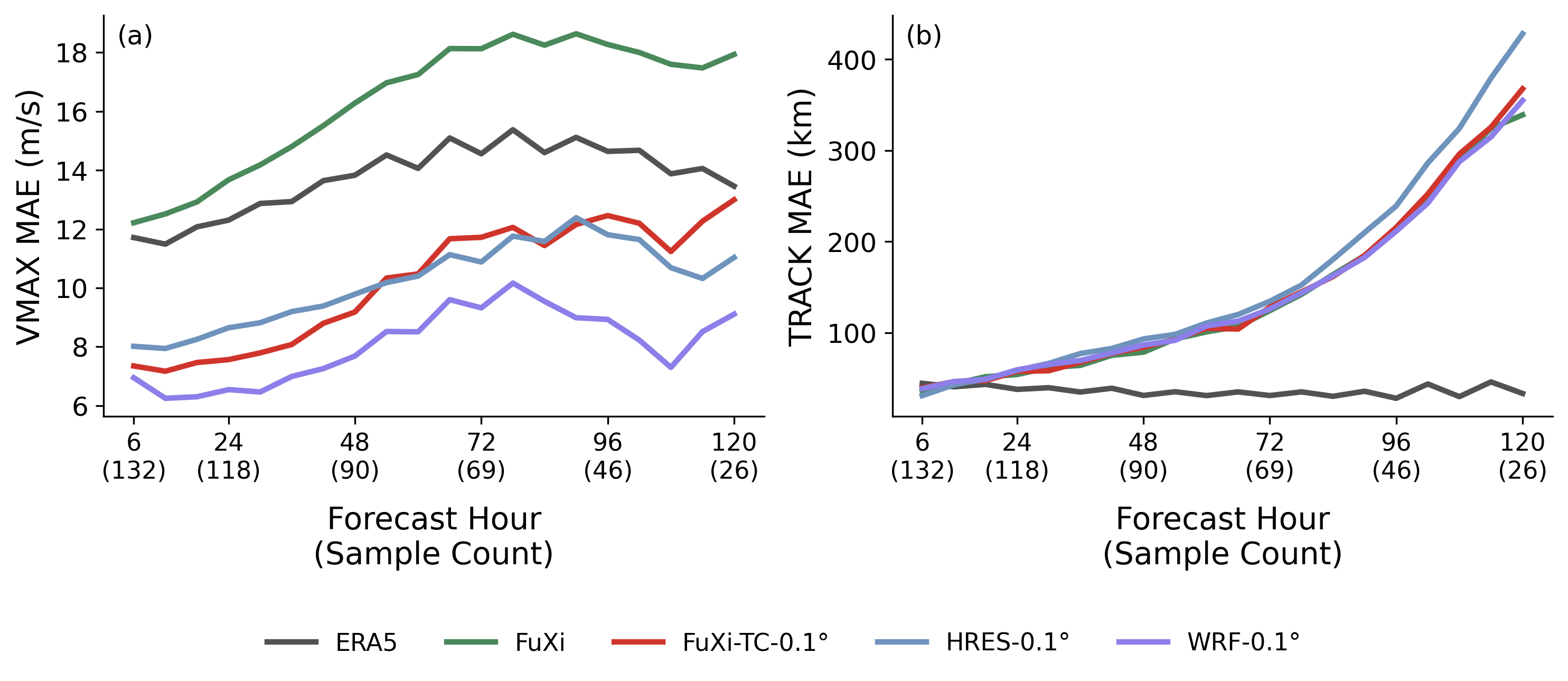}
    \caption{Comparison of TC intensity and track forecasts with IBTrACS dataset as ground truth. Mean absolute error (MAE) of maximum sustained wind and tracks with forecast lead time for ERA5 (black lines), FuXi (green lines), FuXi-TC-\(0.1^\circ\) (red lines), HRES-\(0.1^\circ\) (blue lines) and WRF-\(0.1^\circ\) (purple lines). All the forecast data are evaluated against the IBTrACS dataset. The evaluation covers forecasts for 17 TCs from July to October 2024.}
    \label{statisticmaetrackpacific}
\end{figure}

Evaluation of TC forecasts includes assessments of both track and intensity prediction.
In this study, we have analyzed the forecast performance for 17 TCs ({see Supplementary Table 1}) that occurred during July–October 2024, which represents the typhoon season in the WNP. Fig. \ref{statisticmaetrackpacific} presents a statistical comparison between ECMWF high-resolution deterministic forecast (HRES-\(0.1^\circ\)), FuXi, WRF-\(0.1^\circ\) and FuXi-TC-\(0.1^\circ\) in 5-day forecasts as well as ERA5 reanalysis data against IBTrACS dataset. 
In terms of TC intensity forecasts, FuXi-TC-\(0.1^\circ\) trained using WRF-\(0.1^\circ\) presents a more noticeable improvement over FuXi and ERA5 in terms of the maximum 10-m wind speed (WS10M), as evidenced by lower MAE values for Vmax. Furthermore, FuXi-TC-\(0.1^\circ\) demonstrates overall comparable performance to HRES-\(0.1^\circ\). Specifically, it outperforms HRES-\(0.1^\circ\) during the first 48 hours of the forecast, although it performs marginally worse in the final 12 hours.
Regarding TC track forecasts, FuXi, FuXi-TC-\(0.1^\circ\), and WRF-\(0.1^\circ\) exhibit negligible differences in track errors, and all perform better compared to HRES-\(0.1^\circ\) at lead times of 84-120 hours.

\FloatBarrier

\begin{figure}[htbp]
    \centering
    \includegraphics[width=\linewidth]{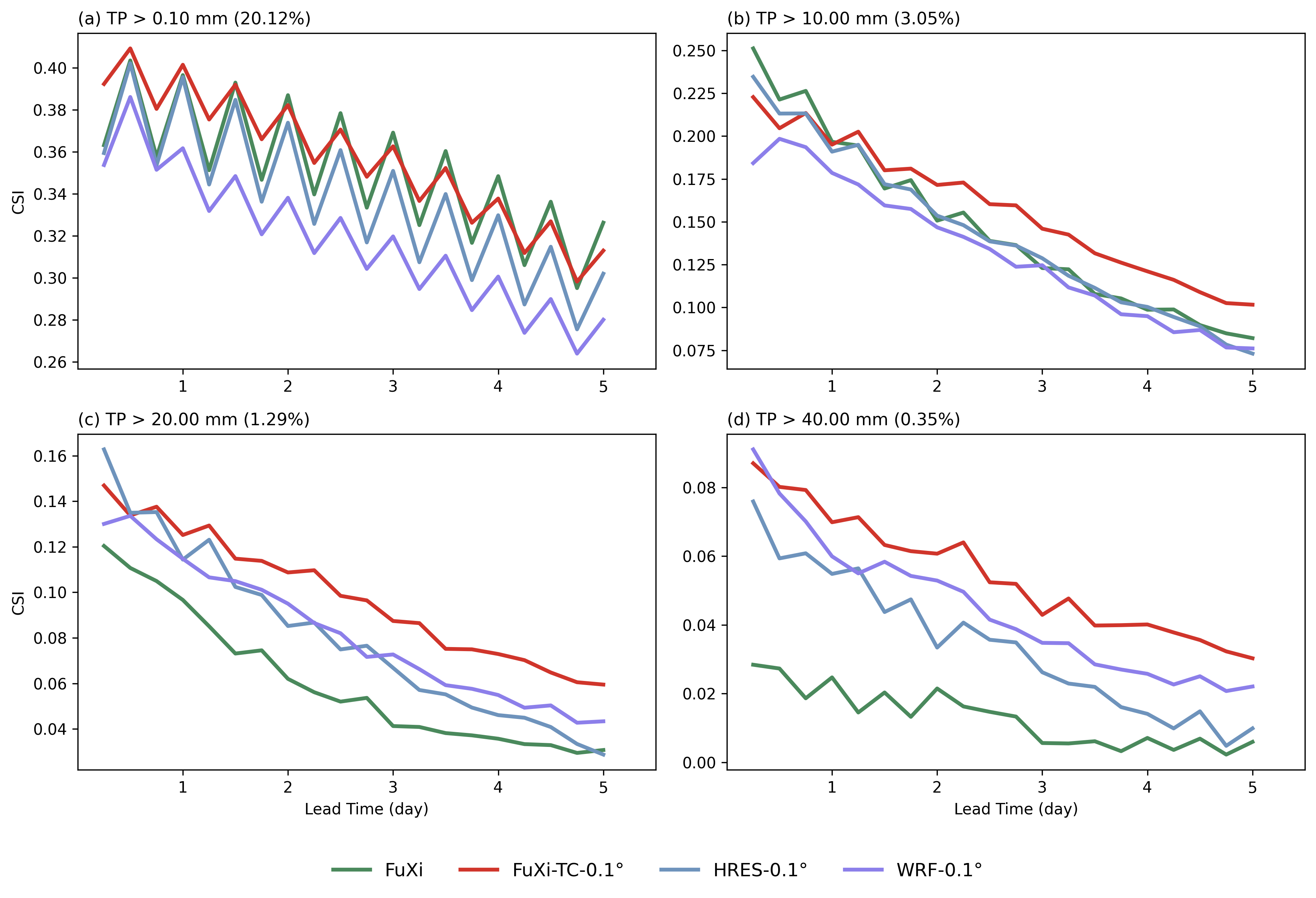}
    \caption{Comparison of CSI for TP. Comparison for the FuXi (green lines), FuXi-TC-\(0.1^\circ\) (red lines), HRES-\(0.1^\circ\) (blue lines) and WRF-\(0.1^\circ\) (purple lines) in predicting 6-hour precipitation for 0.1, 10, 20, 40 mm thresholds in (a-d) respectively. The ratios of extreme cases relative to the entire test set are indicated in parentheses. All the forecast data are evaluated against the Automatic Weather Station observations in period 1 July to 30 September 2024.}
    \label{statistic_csi_precipitation}            
\end{figure}

Beyond track and intensity, accurate rainfall prediction is crucial for assessing the potential impact of a typhoon. Fig. \ref{statistic_csi_precipitation} shows the time series of the Critical Success Index (CSI) of FuXi, FuXi-TC-\(0.1^\circ\), HRES-\(0.1^\circ\) and WRF-\(0.1^\circ\) for 6-hour accumulated total precipitation (TP) evaluated using various threshold values for 5-day forecasts. 
As expected, CSI values decrease with increasing thresholds, indicating greater challenges in predicting more extreme events.
Compared to FuXi forecasts, the physics-based model (including HRES-\(0.1^\circ\) and WRF-\(0.1^\circ\)) maintain a distinct advantage in forecasting heavier precipitation events.
When comparing CSI between FuXi and WRF-\(0.1^\circ\), WRF-\(0.1^\circ\) significantly outperforms FuXi at higher thresholds, however FuXi achieves better CSI scores across a broader range of lower thresholds.
The figure illustrates that FuXi-TC-\(0.1^\circ\) successfully integrates the superior performance of physics-based models in capturing heavy precipitation with the proficiency of FuXi model in predicting general rainfall (indicated by the $\geq$ 0.1 mm threshold).
Consequently, it achieves the highest CSI scores across most lead times and precipitation thresholds, suggesting its superior capabilities in precipitation forecasting.
Notably, the CSI metrics exhibit a sawtooth-like fluctuation, which is particularly evident in the evaluation with a threshold greater than 0.1 mm.
This oscillation stems from differences in valid data availability from AWS observations between 00/12 UTC and 06/18 UTC (see Supplementary Fig. 2), leading to significant fluctuations in forecast false alarms and corresponding variations in the probability of detection (Supplementary Figs. 5 and 6).

\FloatBarrier

\subsection{Deterministic case studies}
To further illustrate the advantages of FuXi-TC-\(0.1^\circ\), we examine Typhoon Bebinca (2024254N10148) and Gaemi (2024201N12133) separately as representative cases of wind speed and precipitation intensity. 

\begin{figure}[htbp]
    \centering
    \includegraphics[width=\linewidth]{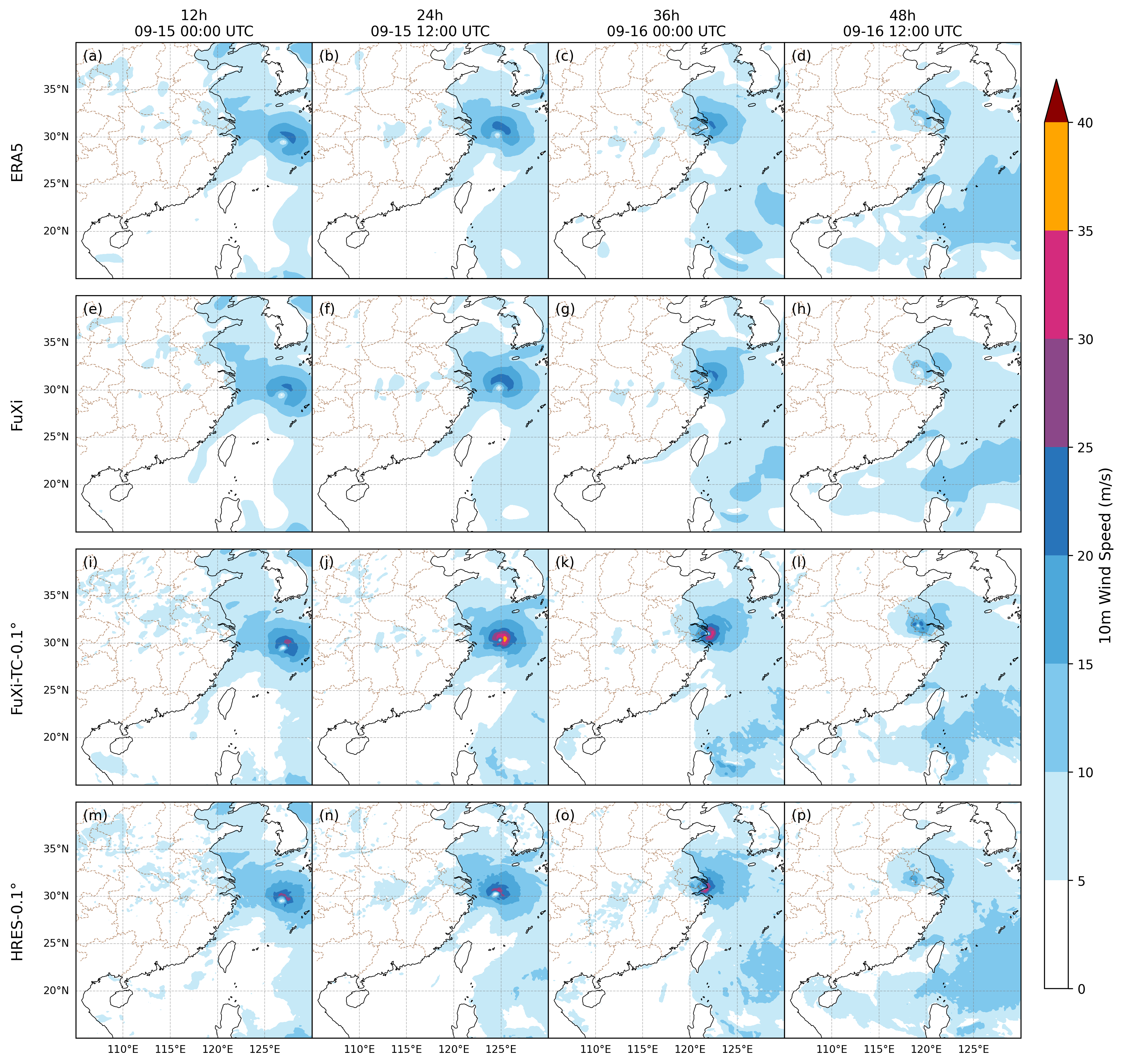}
    \caption{Comparison of WS10M for Typhoon Bebinca forecasts. Comparison of snapshot examples of WS10M among ERA5 (first row), FuXi (second row), FuXi-TC-\(0.1^\circ\) (third row) and HRES-\(0.1^\circ\) (fourth row) for 12 (first column), 24 (second column), 36 (third column) and 48 (fourth row) hours forecasts with the initial time at 12:00 UTC 15 September 2024. }
    \label{bebinca_intensity_case}            
\end{figure}

On 16 September 2024 at 07:30 Beijing time, Bebinca made landfall in Shanghai as a severe typhoon (42 m s$^{-1}$), marking the strongest typhoon to strike Shanghai and Jiangsu since 1949 and causing severe wind and rainfall impacts \cite{wu2025extreme}. 
Fig. \ref{bebinca_intensity_case} shows the forecast initialized at 12:00 UTC on 15 September 2024. This specific time period was selected as an illustrative example because it covers not only the landfall of Typhoon Bebinca but also coincides with the peak intensity of its lifecycle. 
The figure shows that both ERA5 and FuXi dataset underestimate the intensity of Bebinca. FuXi-TC-\(0.1^\circ\) presents stronger wind speeds like HRES-\(0.1^\circ\) also contains finer-scale details in contrast to ERA5 and FuXi. 
The corresponding precipitation forecast for this case is provided in Supplementary Fig. 3, and the typhoon track forecast is shown in Supplementary Fig. 12.

\FloatBarrier

\begin{figure}[htbp]
    \centering
    \includegraphics[width=\linewidth]{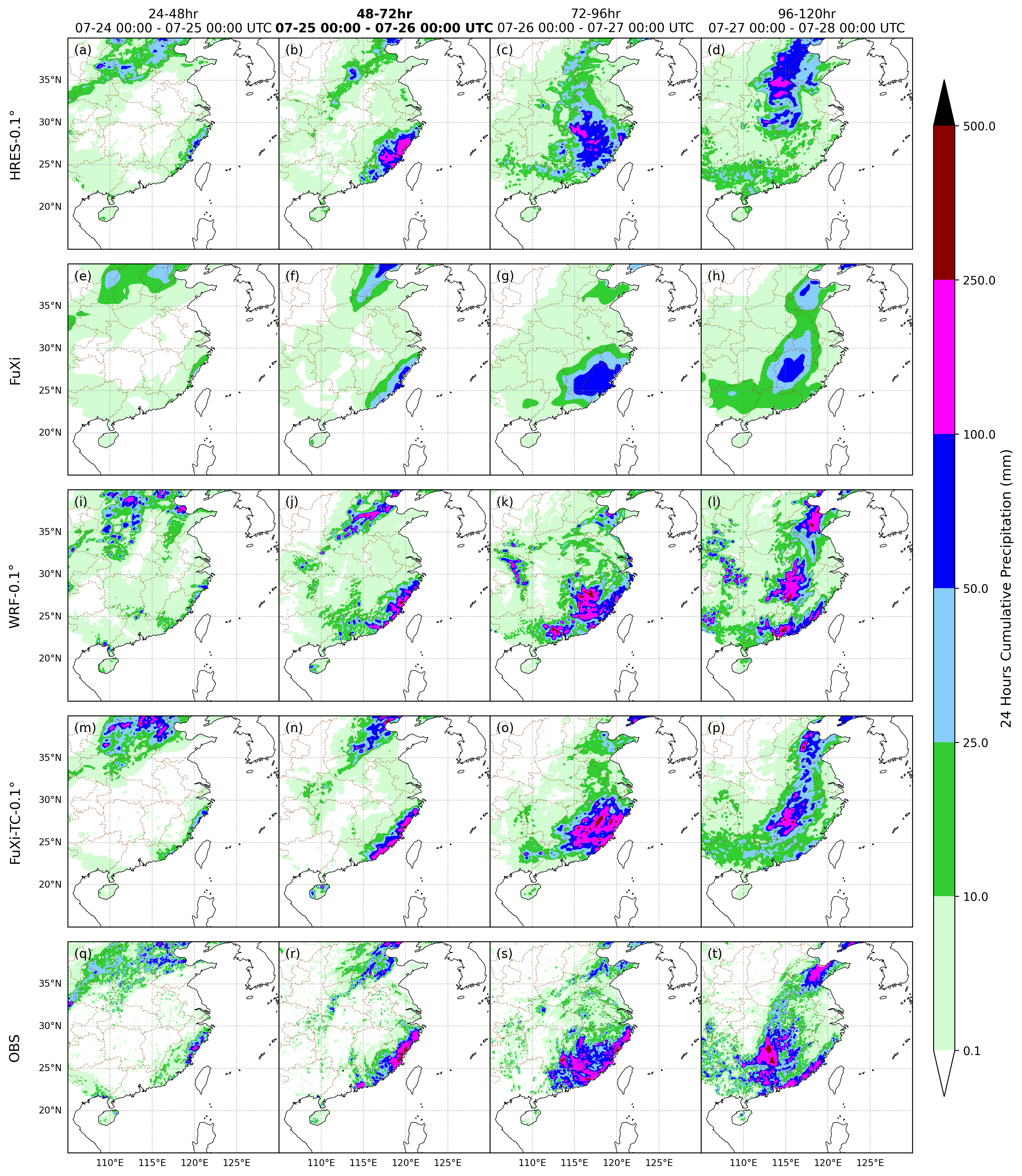}
    \caption{Comparison of TP for Typhoon Gaemi. Comparison of snapshot examples of 24-hour accumulated TP among HRES-\(0.1^\circ\) (first row), FuXi (second row), WRF-\(0.1^\circ\) (third row), FuXi-TC-\(0.1^\circ\) (fourth row) and the Automatic Weather Station observations (fifth row) mainland in China for 24-48 (first column), 48-72 (second column), 72-96 (third column) and 96-120 (fourth row) hours forecasts with the initial time at 00:00 UTC 23 July 2024. }
    \label{gaemi_precip_case}            
\end{figure}

Fig. \ref{gaemi_precip_case} displays the 24-hour accumulated precipitation observations from Automatic Weather Station (AWS) alongside forecasts initialized at 00 UTC on 23 July 2024 from FuXi, FuXi-TC-\(0.1^\circ\), WRF-\(0.1^\circ\) and HRES-\(0.1^\circ\), over the periods of 24-48, 48-72, 72-96 and 96-120 hours. 
This case study was selected to exemplify the models' performance during Super Typhoon Gaemi, which triggered record-breaking extreme rainfall since 1961 as it made landfall in Fujian province \cite{wu2025extreme}. 
For clarity, precipitation values are shown exclusively for mainland China, with oceanic and external landmasses masked. 
The figure reveals \(0.1^\circ\)-resolution models generate more detailed rainfall forecasts compared to FuXi (\(0.25^\circ\)). 
Regarding the prediction of the heavy precipitation areas, FuXi-based forecasts align better with the observation, compared to HRES-\(0.1^\circ\), especially at lead times of 4-5 days. 
In terms of rainfall intensity, FuXi evidently underestimates the peak values relative to observations. In contrast, FuXi-TC-\(0.1^\circ\) yields a more accurate intensity forecast that agrees with observations and is comparable to WRF-\(0.1^\circ\).
Furthermore, while WRF-\(0.1^\circ\) produces more false alarms in many non-precipitation areas compared to FuXi, FuXi-TC-\(0.1^\circ\) demonstrates higher accuracy in these rain-free regions.
The corresponding track forecast for this case is provided in  Supplementary Fig. 12.

\FloatBarrier

\subsection{Ensemble forecast comparison}

\begin{figure}[htbp]
    \centering
    \includegraphics[width=\linewidth]{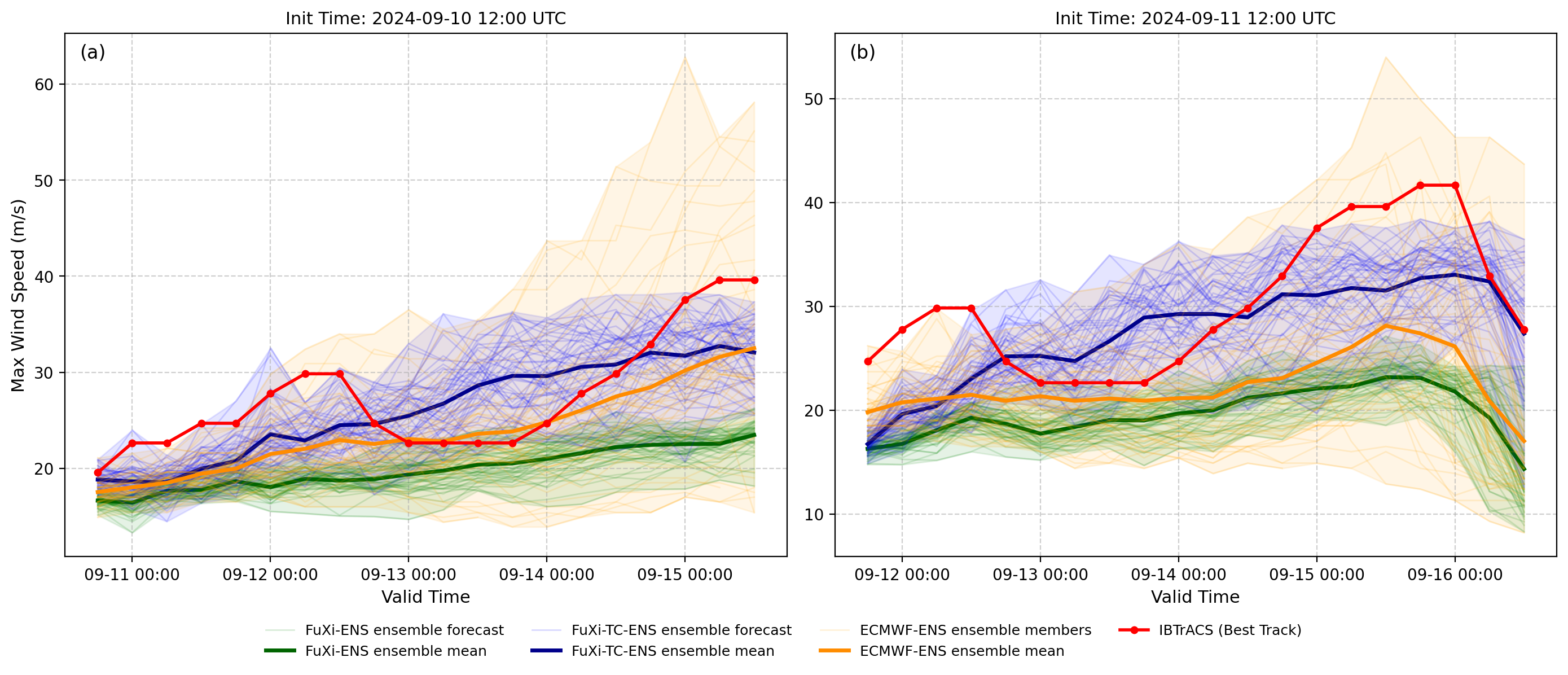}
    \caption{Comparison of maximum sustained wind speed forecasts for Typhoon Bebinca. Comparison the maximum sustained wind speed from IBTrACS (red line) and ensemble forecasts generated by FuXi-ENS (green line), FuXi-TC-ENS (blue line) and ECMWF-ENS (orange line) with the initial forecasting time at 12:00 UTC 10 September and 12:00 UTC 11 September 2024.}
    \label{bebinca_ens}            
\end{figure}

Unlike deterministic forecasts, ensemble forecasts offer the distinct advantage of providing probabilistic guidance. While FuXi-ENS demonstrates superior performance over ECMWF-ENS with a comprehensive evaluation in key forecast metrics such as Continuous Ranked Probability Score (CRPS) and Brier Score (BS), challenges regarding ensemble under-dispersion and insufficient uncertainty quantification for extreme phenomena persist \cite{zhong2024fuxiens}. 
Following the FuXi-TC framework, a diffusion generative module is utilized to refine each FuXi-ENS member, yielding the enhanced ensemble system designated as FuXi-TC-ENS, which is then compared against ECMWF-ENS and FuXi-ENS in terms of TC intensity forecasts using 51 ensemble members.
Fig. \ref{bebinca_ens} illustrates temporal evolution of max sustained wind speed forecasts for Typhoon Bebinca. 
FuXi-ENS exhibits a notable underestimation of maximum sustained 10-m wind speeds compared to IBTrACS data. Specifically, the model's simulated peak wind speeds are limited to approximately 20 m/s, whereas observed values exceed 40 m/s. Furthermore, FuXi-ENS displays insufficient ensemble spread, which is likely due to the limited projection of initial perturbations and the model's inability to sufficiently amplify forecast perturbations over time \cite{zhong2024fuxi-exteme}. 
Unlike simple additive noise, the diffusion-based refinement module enhances each ensemble forecast by superimposing generative perturbations onto FuXi-ENS. Consequently, FuXi-TC-ENS improves peak intensity estimation and exhibits a more realistic increase in ensemble spread with lead time without degrading track accuracy.
The corresponding ensemble track forecasts for this case are provided in Supplementary Fig. 4.

\FloatBarrier

\subsection{Generalization to the NA Basin}
\begin{figure}[htbp]
    \centering
    \includegraphics[width=\linewidth]{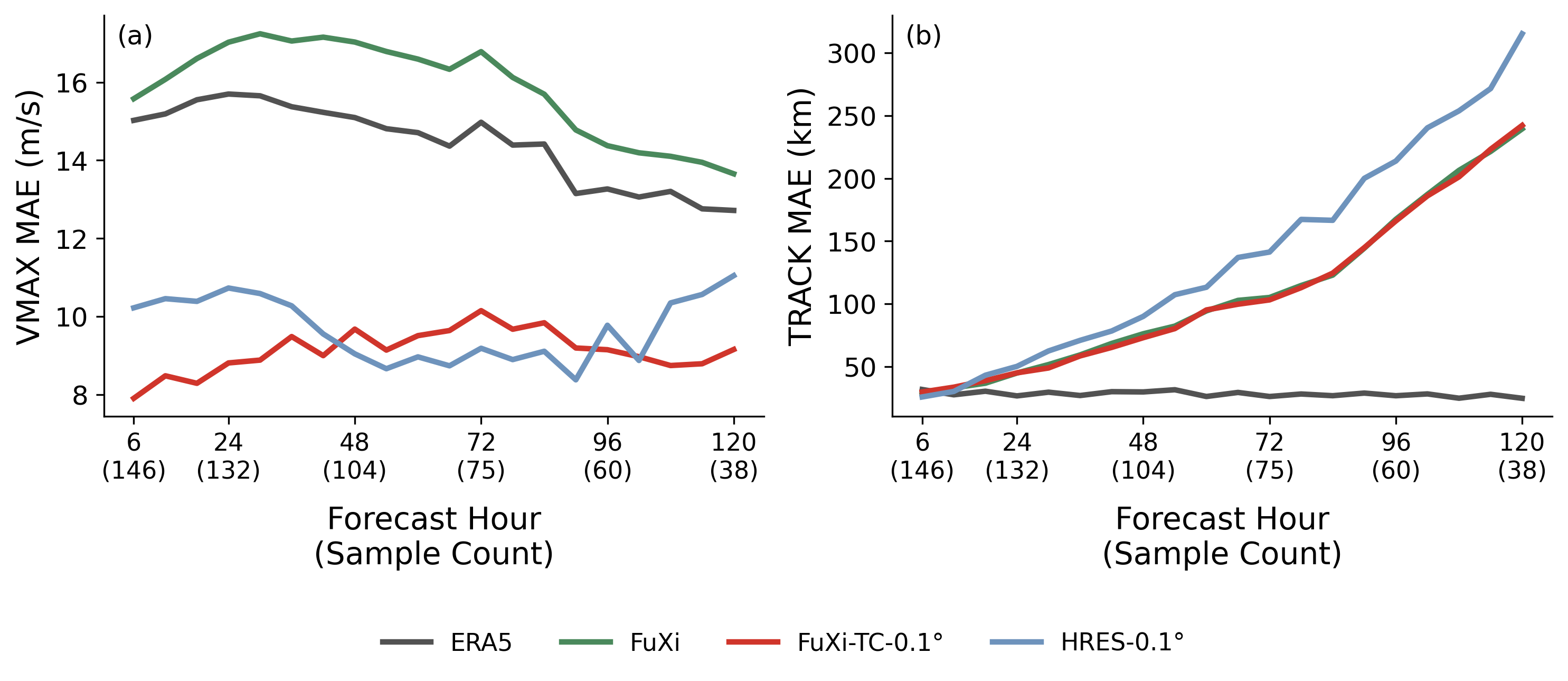}
    \caption{Comparison of TC intensity and track forecasts in North Atlantic. MAE of maximum sustained wind and tracks with forecast lead time for ERA5 (black lines), FuXi (green lines), FuXi-TC-\(0.1^\circ\) (red lines) and HRES-\(0.1^\circ\) (blue lines). All the forecast data are evaluated against the IBTrACS dataset. The evaluation covers forecasts for 15 TCs from June to November 2024.}
    \label{statistic_mae_track_atlantic}            
\end{figure}

To evaluate the generalization capability of FuXi-TC, we directly applied the model trained on the WNP to the NA region. Constrained by the fixed spatial dimensions of the WNP training domain, we partitioned the NA region into two overlapping sub-domains: a western segment and an eastern segment (see Supplementary Fig. 11).
We then reconstructed the full-field forecast for the entire NA basin by stitching these sub-domains together, where predictions in the overlapping areas were averaged to ensure spatial continuity.
By analyzing 15 TCs ({see Supplementary Table 1}) that occurred from June to November 2024, we present a statistical comparison of intensity and track forecasting performance among HRES-\(0.1^\circ\), ERA5, FuXi and FuXi-TC-\(0.1^\circ\) in Fig.~\ref{statistic_mae_track_atlantic}.
In terms of TC intensity forecasting, FuXi-TC-\(0.1^\circ\) still significantly outperforms FuXi and ERA5 in predicting the maximum WS10M. Moreover, FuXi-TC-\(0.1^\circ\) delivers superior performance to HRES-\(0.1^\circ\) within the first 36 hours and last 12 hours of forecasting, yet it underperforms HRES-\(0.1^\circ\) in the mid-term forecasting period.
Regarding TC track forecasting, compared with HRES-\(0.1^\circ\), the results of FuXi-based models (including FuXi and FuXi-TC-\(0.1^\circ\)) exhibit better performance during the 48–120h forecasting window. To further illustrate this generalization capability, detailed case studies of North Atlantic hurricanes Beryl and Helene are provided (Supplementary Figs. 9 and 10). These cases demonstrate that FuXi-TC-\(0.1^\circ\) successfully reconstructs intense, fine-scale inner-core wind structures, even in unseen geographical basins.

\FloatBarrier

\subsection{Computational efficiency comparison}

Compared with traditional NWP models, a primary advantage of deep learning approaches is their drastically improved computational efficiency and reduced resource demands, particularly for large-domain and high-resolution forecasts. As detailed in Table~\ref{tab:runtime_comparison}, the WRF model at \(0.1^\circ\) resolution requires approximately 11.5 hours on 32 CPUs to complete a single 120-hour forecast under typical operational configurations. In contrast, FuXi-TC-\(0.1^\circ\) generates a forecast with equivalent lead time and resolution in just 2 minutes on a single GPU, corresponding to a speedup of over two orders of magnitude. This efficiency enables rapid update cycles, facilitates large-scale ensemble forecasting, and significantly lowers the computational cost of high-resolution TC prediction, making FuXi-TC-\(0.1^\circ\) feasible for real-time operational applications.

\setlength{\tabcolsep}{2pt}
\begin{table}[h]
\caption{
Resource requirements and runtime per task.
}
\centering
\vspace{-5pt}
\begin{tabular}{lccr}
\toprule
\textbf{Task} & \textbf{Resource} & \textbf{Quantity} & \textbf{Runtime (per forecast)} \\
\midrule
WRF-\(0.1^\circ\) & Intel Xeon Platinum 8369B CPU @ 2.90GHz & 32 &  11.5 hours \\
FuXi-TC-\(0.1^\circ\) & NVIDIA A100 GPU & 1 & 2 minutes \\
\bottomrule
\end{tabular}
\label{tab:runtime_comparison}
\end{table}
\vspace{-5pt}

\section{Discussion}

Accurate prediction of TC intensity remains a significant challenge for both NWP and emerging AI forecasting systems.
While recent AI–based weather forecasting models have achieved substantial improvements in track prediction, they often underestimate TC maximum wind speeds and associated precipitation.
A major contributing factor is the intrinsic smoothing present in commonly used training datasets, particularly ERA5, which often suppresses extreme wind speeds and precipitation and blurs fine-scale structures associated with intense storms.
In this study, we introduce FuXi-TC, a diffusion-based generative forecasting framework designed to mitigate these limitations.
The model improves predictions of WS10M, TP and MSL while preserving the high accuracy of underlying FuXi-2.0 track forecasts.
By recovering sharper spatial structures and strong extremes, the approach provides a computationally efficient alternative to traditional dynamical downscaling methods.
Although WRF simulations are not bias-free, they provide dynamically consistent high-resolution structures that are absent in reanalysis datasets, making them suitable training targets for mesoscale refinement.

To address the smoothing bias prevalent in AI predictions, previous studies have primarily resorted to directly dynamical downscaling using NWP models. Despite their efficacy in recovering small-scale details, the prohibitive computational expense of such methods renders them difficult to scale for large-ensemble configurations or operational use.
Beyond computational efficiency, a key advantage of the FuXi-TC framework is its robust transferability across both base forecasting models and geographical domains. Although trained exclusively on deterministic FuXi-2.0 forecasts over the WNP, the generative module can be directly applied to other systems and regions. For instance, while the original FuXi-ENS achieves strong general forecast metrics, it suffers from significant under-dispersion and tends to underestimate the intensity of extreme events. Applying FuXi-TC to refine individual ensemble members of FuXi-ENS produces the FuXi-TC-ENS system, which increases ensemble spread without degrading track consistency and improves the representation of high-impact extremes. 
Unlike recent hybrid ensemble methods that rely on computationally expensive full physics-based models to couple with machine-learned flows \cite{polichtchouk2026hybridensembleforecastingcombining}, our framework achieves this enhancement without incurring massive computational costs. Its lightweight and highly transferable nature allows it to be seamlessly extended to large-ensemble scenarios.

Additionally, the model exhibits strong zero-shot geographical generalization. Despite being trained on the WNP basin, FuXi-TC performs robustly in the NA basin without fine-tuning, suggesting that the model may capture transferable physical structures associated with TC evolution rather than simply reproducing region-specific patterns.
More broadly, this work highlights the potential of combining AI models with on NWP-derived data to extend the capabilities of AI weather forecasting.
Recent analyses have suggested that the supply of high-quality human-generated training data for AI systems is becoming increasingly limited \cite{villalobos2024position}.
Synthetic data generated by models can partially address this constraint, but over-reliance on such data may degrade model performance and diminish creativity and introduce bias \cite{shumailov2024ai,gerstgrasser2024model}.
In weather forecasting, the success of AI models has relied on extensive, high-quality, open-source datasets, such as ERA5, which span multiple decades.
However, these datasets are finite, and the ERA5 archive has effectively been exhausted.
Therefore, NWP models provide a complementary source of synthetic data that can augment training datasets and expose AI models to rare or extreme weather scenarios, e.g., TC \cite{bauer2024if}.
Indeed, ERA5 itself can be viewed as a synthetic dataset, produced by ECMWF through 4D-Var data assimilation combined with CY41R2 model forecasts.

Beyond advancing TC forecasts, we demonstrate the broader significance of leveraging NWP-derived datasets to push the performance limits of AI-based weather forecast.
However, a critical limitation of this framework is that FuXi-TC's performance upper bound may ultimately be locked by the systematic biases inherent in NWP physical parameterizations. To break through this theoretical ceiling, our future work will focus on directly integrating real-world observations to correct these NWP-induced biases.

\section{Method}

\subsection{Dataset}
\subsubsection{Tropical cyclone dataset}
To evaluate TC intensity and trajectory forecasts, we employ IBTrACS, a globally harmonized TC database maintained by NOAA National Centers for Environmental Information (NCEI), which integrates best-track records from multiple meteorological agencies \cite{knapp2010international, kenneth2019international}.
The dataset includes TC center position, maximum sustained surface wind speed, and minimum central pressure (hPa).
For validation against IBTrACS, we utilized the CMA agency best track data in the WNP and USA agency best track data in the NA respectively (see Supplementary Fig. 11 for the spatial domains).

To derive the forecasted TC tracks and intensities, we employ a modified version of the Aurora tracker \cite{bodnar2025foundation}. The tracker identifies the TC center by detecting local minima in the MSL field—with 700 hPa geopotential height (Z700) as a fallback. Once the center is located, the forecasted intensity is determined by extracting the minimum MSL and maximum WS10M within a specific radius. Following the configuration of FuXi-Extreme \cite{zhong2024fuxi-exteme}, we incorporate physical termination criteria to prevent spurious tracking artifacts when the cyclone dissipates. Specifically, tracking is halted if the maximum 850 hPa relative vorticity and maximum WS10M falls below specific threshold values. This approach ensures alignment with standard evaluation protocols for TC forecasting.

For comparison with the ECMWF High-Resolution Forecast including both deterministic (HRES) and ensemble forecasts (ENS) efficiently, we acquired TC track data from the THORPEX Interactive Grand Global Ensemble (TIGGE) archive \cite{bougeault2010thorpex,swinbank2016tigge}.
This dataset incorporates tracks generated by the operational ECMWF tracker and is distributed in XML format, where the ECMWF HRES and ENS tracks are explicitly identified by the "Forecast" and "ensembleForecast" tags, respectively.

\subsubsection{FuXi-2.0 dataset} 
Serving as the primary source of training dataset, FuXi-2.0 represents a significant advancement over its predecessor FuXi-1.0 \cite{chen2023fuxi}, serving as an AI-based medium-range weather forecasting model with a coupled ocean–atmosphere structure.
A key innovation is FuXi-2.0 employs a dual-model framework where a transformer-based interpolator efficiently derives continuous 1-hourly forecasts from 6-hourly predictions, thereby minimizing iterative steps and ensuring temporal coherence.

Furthermore, by explicitly modeling ocean feedback processes like sea surface heat fluxes, FuXi-2.0 improves the representation of air–sea interactions, which is critical for capturing the key features of TCs \cite{zhong2024fuxi2.0}.
The model is trained and driven using the ERA5 reanalysis dataset, covering the period from 2012 to 2017.

This study uses FuXi-2.0 fields as the main source of global atmospheric initial and boundary conditions for the WRF model, with soil variables and sea surface temperature (SST) provided by ERA5 (A summary of variable deﬁnitions can be referred to in Table ~\ref{tab:wrf_variables}). The following surface variables are used: mean sea level pressure (MSL), 2-meter temperature (T2M), 2-meter dew point temperature (D2M), surface pressure (SP), and 10-m zonal and meridional winds (U10M and V10M).
Upper-air variables on 13 standard pressure levels (50, 100, 150, 200, 250, 300, 400, 500, 600, 700, 850, 925, and 1000 hPa) include geopotential (Z), temperature (T), zonal wind (U), meridional wind (V), and relative humidity (RH), enabling a comprehensive representation of the tropospheric state. To ensure alignment with the variable requirements of the WRF model, RH is derived from specific humidity (Q), pressure (P), and temperature (T) using standard thermodynamic conversions.
First, the actual vapor pressure is calculated as follows \cite{wallace2006atmospheric}:
\begin{equation}
    e = \frac{\mathrm{Q} \times \mathrm{P}}{0.622 + 0.378 \mathrm{Q}}
\end{equation}
where \( \mathrm{Q} \)  is the specific humidity in kg/kg, and \( \mathrm{P} \)  is the atmospheric pressure in hPa. Next, the saturation vapor pressure \(e_s\) (in hPa) is estimated using the empirical formula proposed by Bolton \cite{bolton1980computation}:
\begin{equation}
    e_s = 6.112 \exp\left( \frac{17.67\,\mathrm{T_c}}{\mathrm{T_c} + 243.5} \right)
\end{equation}
where \( \mathrm{T_c} \) is the temperature in °C (\( \mathrm{T_c} = \mathrm{T} - 273.15 \)). Finally, the relative humidity is calculated as the ratio of actual to saturation vapor pressure, capped at 100\%:
\begin{equation}
    \mathrm{RH} = \frac{e}{e_s} \times 100\%
\end{equation}

\begin{table}[htbp]
\centering
\caption{Summary of initial and boundary condition variables for the WRF model.}
\label{tab:wrf_variables}
\renewcommand{\arraystretch}{1.2} 
\begin{tabular}{@{} p{3cm} p{6.5cm} p{2.5cm} @{}}
\toprule
\textbf{Category} & \textbf{Variables} & \textbf{Source} \\ \midrule
\multirow{7}{*}{\textbf{single level}} 
 & mean sea level pressure & FuXi-2.0 \\
 & surface pressure & FuXi-2.0 \\
 & 2-m temperature & FuXi-2.0 \\
 & 2-m dew point & FuXi-2.0 \\
 & 10-m zonal/meridional winds & FuXi-2.0 \\ 
 & soil variables & ERA5 \\
 & sea surface temperature & ERA5 \\ \midrule
\multirow{5}{*}{\textbf{\begin{tabular}[c]{@{}l@{}}upper-air\\ (13 pressure levels)\textsuperscript{a}\end{tabular}}} 
 & geopotential & FuXi-2.0 \\
 & temperature & FuXi-2.0 \\
 & zonal/meridional wind & FuXi-2.0 \\
 & specific humidity & FuXi-2.0 \\
 & relative humidity & FuXi-2.0(Derived) \\ \bottomrule
\multicolumn{3}{l}{\footnotesize \textsuperscript{a} 50, 100, 150, 200, 250, 300, 400, 500, 600, 700, 850, 925, and 1000 hPa.} \\
\end{tabular}
\end{table}

\subsubsection{WRF dataset}
High-resolution training targets are created using simulations from the WRF model (detailed parameter configurations are provided in Section \ref{sec:wrf_conf}).
In contrast to previous hybrid approaches \cite{liu2024hybrid, xu2025exploring, niu2025improving, husain2025leveraging} that initialize with Pangu-weather or Graphcast to provide initial and boundary conditions, the simulations in this study were driven primarily by FuXi-2.0 forecasts.
This configuration preserves structural consistency between the large-scale input forecasts and the high-resolution target fields used for model training.

The WRF simulations were configured on a regular latitude-longitude grid with a horizontal resolution of \(0.1^\circ\), comprising 592×496 grid points in longitude and latitude. In the vertical direction, the model atmosphere was discretized into 56 eta levels, extending to a model top pressure of 50 hPa.
The downscaled fields provide spatially detailed representations of TC structure, including strong wind gradients and precipitation patterns that are not resolved in coarser and smoother AI forecasts.
We select a set of key surface meteorological variables, including T2M, U10M, V10M, WS10M, MSL, and 6-hourly accumulated TP.
Upper-air variables are extracted for pressure levels of 200, 300, 500, 700, and 850 hPa, including T, Q, U and V.
Upper-air variables were vertically interpolated from model levels to the specified pressure levels using linear interpolation.
These variables are integrated into the FuXi-TC framework to effectively capture the feature of TC.

\subsubsection{Training data preprocessing}
Preprocessing was applied separately to different variable types. For non-precipitation variables, Z-score normalization is applied using the mean and standard deviation computed from the training dataset. For TP, which exhibits a highly skewed, long-tailed distribution, we employ a logarithmic transformation to mitigate the dominance of extreme values and yield a more Gaussian-like distribution \cite{kuhn2013applied,shrestha2019evaluation}. Specifically, the transformation is defined as:
\begin{equation}
\tilde{x} = \ln(x + 1)
\end{equation}
where \(x\) is the original 6-hourly accumulated precipitation, and \(\tilde{x}\) is the transformed value. The constant \(1\) is added to handle zero precipitation values.

\subsection{FuXi-TC model structure}

\begin{figure}[htbp]
    \centering
    \includegraphics[width=\linewidth]{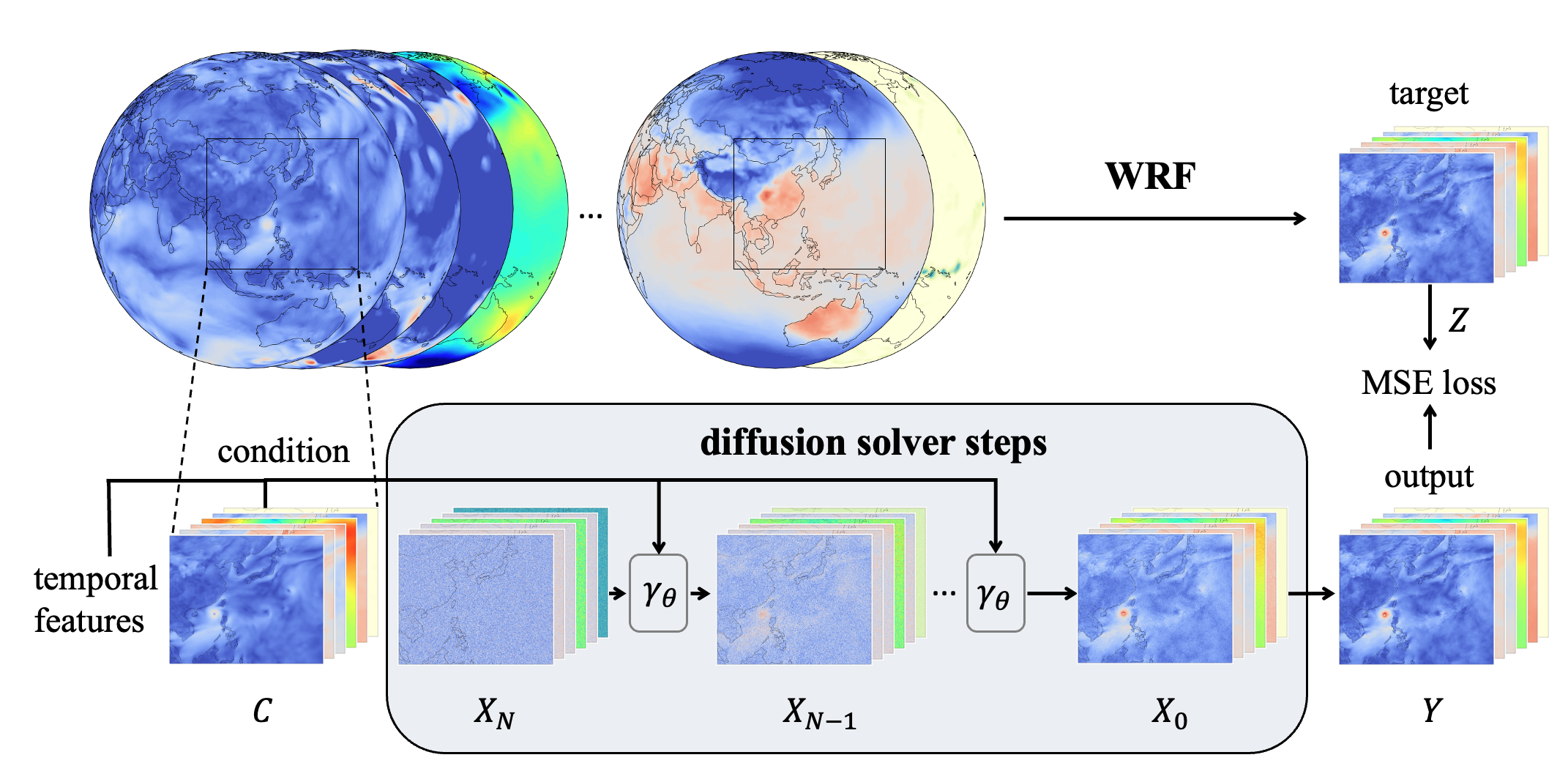}
    \caption{Schematic diagram of the structures of the FuXi-TC framework. Coarse-resolution regional subsets are extracted from the global FuXi-2.0 forecasts to serve as the primary conditioning inputs. The generative refinement module, based on DDPM iteratively denoises random noise through a UNet-based solver. The denoising process is jointly conditioned on the meteorological fields and encoded temporal features. During training, the model is optimized using a MSE loss to map the generated high-resolution forecasts to physically consistent WRF simulations. }
    \label{fig:fuxi_tc_framework}
\end{figure}

As illustrated in Fig.\ref{fig:fuxi_tc_framework}, the FuXi-TC framework employs a hybrid architecture that integrates global large-scale meteorological guidance with a generative refinement module. 
This design is inspired by the recent success of the FuXi-Extreme model \cite{zhong2024fuxi-exteme}, which demonstrated that coupling autoregressive forecasting with diffusion models significantly improves the representation of extreme weather events. 
Specifically, the framework extracts regional subsets from the FuXi-2.0 global forecast dataset (see regional domain in Supplementary Fig. 11). These coarse-resolution forecasts serve as the primary conditioning inputs for the downscaling process.
In addition to these meteorological fields, the generative process is further conditioned on the diffusion noise level and various temporal features, including the forecast step, day of the year, and hour of the day, are encoded using sine/cosine transformations to accurately guide the denoising trajectory.

The final high-resolution forecasts are achieved via a Denoising Diffusion Probabilistic Model (DDPM) \cite{ho2020denoising,sohl2015deep}, which is trained to map the coarse FuXi-2.0 priors to physically consistent WRF simulations. 

\FloatBarrier

\subsection{Denoising diffusion probabilistic model (DDPM)}
The DDPM serves as the generative backbone of FuXi-TC, which treats the downscaling and bias-correction task as a conditional generation problem, similar to image super-resolution strategies employed in computer vision \cite{saharia2022image}.

We employ the UNet \cite{Ronneberger2015} architecture from the Diffusers\footnote{https://github.com/huggingface/diffusers} library to build our DDPM model, which includes three downsampling layers and three upsampling layers.
Each layer applies 2D convolutions with $3 \times 3$ kernels, SiLU activation functions \cite{ramachandran2017searching} and group normalization\cite{wu2018group}, and the number of channels are 256, 512 and 1024 for the first, second, and third layers respectively.
At the bottleneck of the network, 12 global self-attention layers \cite{vaswani2017attention} are integrated to enhance the capture of long-range spatial dependencies, which is crucial for restoring TC fine structures such as eye walls and spiral rain bands.

Formally, the diffusion process consists of a forward noise-injection process and a reverse denoising process. In the forward process, random Gaussian noise $\epsilon$ is gradually added to the ground-truth WRF data ($Z$) over $t$ steps ($t \in [1, T]$) according to a variance schedule $\beta_t$~\cite{nichol2021improved}. The noisy sample $X_t$ at step $t$ is expressed as follows,
\begin{equation}
X_t = \sqrt{\bar{\alpha}_t}X_0 + \sqrt{1-\bar{\alpha}_t}\epsilon, \quad \epsilon \sim \mathcal{N}(0, I)
\end{equation}
where \(\alpha_t = 1 - \beta_t\) and \(\bar{\alpha}_t\) is the cumulative product of \(\alpha\), $\bar{\alpha}_t=\prod_{i=1}^t\alpha_i$, $\alpha_i=1-\beta_i$. 

In the reverse process, the model learns to predict the added noise: \(\epsilon_\theta(X_t, t, C)\) with respect to the input condition $C$. The posterior distribution $\gamma_\theta(X_{t-1}|X_t,C)$ is assumed to be Gaussian, whose mean $\mu$ parameterized by the UNet model with parameters $\theta$:

\begin{equation}
\mu_\theta(X_t,t,C) = \frac{1}{\sqrt{\alpha_t}}\left(X_t-\frac{1-\alpha_t}{\sqrt{1-\bar{\alpha}_t}}\epsilon_\theta(X_t,t,C)\right)
\label{eq_mu}
\end{equation}

By predicting the noise, the model learn to reconstruct the original data from the sampled noise $X_T$ via iterative denoising.

The loss function of the DDPM model adopts mean squared error (MSE), which minimizes the difference between the noise added in the forward process and the noise predicted by the model.
\begin{equation}
L = \mathbb{E}_{t, X_0, \epsilon, C} \left[ ||\epsilon - \epsilon_\theta(X_t, t, C)||^2 \right]
\end{equation}
During inference, sampling was performed using 5 denoising steps with a linear beta schedule. This configuration provides a balance between computational efficiency and reconstruction fidelity.

\subsection{FuXi-TC model training}

The FuXi-TC model is trained based on the PyTorch \cite{paszke2019pytorch} framework, and the training process only updates the parameters of the DDPM model, while the parameters of FuXi remain frozen to ensure the stability of the base forecast.
The training is carried out on 2 Nvidia A100 GPUs, with a batch size of 2 per GPU and a total of 60000 iterations.
We use the AdamW \cite{loshchilov2017decoupled} optimizer for training, with the parameters set as $\beta_{1}=0.9$, $\beta_{2}=0.95$.
We select the initial learning rate of $2.5\times$10$^{-4}$ to balance the convergence speed and stability, and use the weight decay coefficient 0.1 to suppress overfitting.

\subsection{WRF model configuration}
\label{sec:wrf_conf}
We employ WRF version 4.3 \cite{jensen2021description} to conduct regional simulations over the WNP (see Supplementary Fig. 11). The model is initialized twice daily at 00 and 12 UTC, with forecasts extending up to 120 hours. 
To facilitate numerical integration and subsequent post-processing training framework, a latitude–longitude projection is adopted.
The simulation domain spans \(100^\circ\)E--\(159.1^\circ\)E and \(0.5^\circ\)N--\(50^\circ\)N, a coverage sufficient to capture the majority of TC tracks in the WNP. This corresponds to a horizontal grid dimension of 592 × 496 points with \(0.1^\circ\) horizontal interval. In the vertical direction, the model top is set at 50 hPa, with 56 vertical levels defined by eta coordinates.

We evaluated the model's sensitivity to horizontal resolution by comparing simulations at \(0.25^\circ\) and \(0.1^\circ\) using identical physical configurations. Supplementary Fig. 1 illustrates the comparative performance regarding TC track and intensity forecasting. The results indicate that while track forecast accuracy remains similar between the two resolutions, the WRF-\(0.1^\circ\) configuration demonstrates significantly superior performance in predicting TC intensity compared to WRF-\(0.25^\circ\). Consequently, the \(0.1^\circ\) horizontal resolution was selected for this study. The WRF outputs at this selected resolution are generated for specific periods in 2023 and 2024. The WRF outputs at this selected resolution are generated for specific periods. We utilize data from April 1 to October 31 of 2023 for model training and reserve data from July 1 to October 31 of 2024 for independent testing.

Both physical parameterization and spectral nudging schemes follow those used in the Shanghai Typhoon Model (SHTM) \cite{niu2025machine}. Specifically, the model employs the Thompson microphysics scheme \cite{thompson2004explicit}, the multi‐scale Kain–Fritsch cumulus scheme \cite{zheng2016improving}, the RRTMG scheme for longwave radiation \cite{iacono2008radiative}, the unified Noah land‐surface model \cite{ek2003implementation}, and the Yonsei University planetary boundary layer scheme \cite{hu2013evaluation}.

Spectral nudging is applied to maintain consistency between the WRF simulations and FuXi’s large‐scale circulation while allowing the model to develop its own mesoscale structures. The nudging configuration is optimized in terms of nudged variables, vertical application, horizontal wavelength cutoff, and relaxation time. Only the zonal and meridional wind components and virtual temperature are nudged, while specific humidity is excluded to prevent degradation of TC intensity forecasts \cite{husain2025leveraging}.

Vertically, nudging is applied only above 850 hPa, excluding the planetary boundary layer to avoid adverse effects arising from FuXi’s limited vertical detail near the surface and from differences in surface forcing over complex terrain. A height‐dependent weighting reduces the nudging influence toward the surface, thereby preserving boundary‐layer processes\cite{husain2025leveraging}.

In spectral space, only large‐scale features exceeding a specified wavelength threshold are constrained, ensuring that synoptic‐scale patterns from FuXi are retained while WRF generates smaller‐scale variability\cite{niu2025machine}. A relatively weak relaxation coefficient provides gentle constraints, enabling the model to align with FuXi’s large‐scale fields without suppressing mesoscale dynamics. Nudging is activated from the start of the integration and maintained throughout the forecast period to ensure persistent large‐scale alignment.

\subsection{Evaluation method}
\subsubsection{TC tracking method}

In our study, we adopt a modified version of Aurora tracker \cite{bodnar2025foundation} to all forecast model's performance including FuXi, FuXi-TC, FuXi-ENS and FuXi-TC-ENS. Initially, we outline the Aurora tracker, then detail our modifications and the justification for these changes.

The primary function of the Aurora tracker is to identifies TC center positions by detecting local minima in predicted MSL fields, with Z700 employed as a fallback variable when MSL-based detection fails when plagued by multiple local minima. The tracker is initialised at the observed TC position from IBTrACS best track data (China Meteorological Administration agency for WNP and United States agency for NA) at the analysis time. For each subsequent 6-hourly forecast lead time, a first-guess position is estimated by linear extrapolation of the preceding tracked positions (up to the most recent eight estimates, corresponding to 48 h). 
To locate the TC center, the tracker searches for the local minimum of the Gaussian-smoothed MSL field closest to the first-guess position. Within a search box of $x^\circ × x^\circ$ centered on the first-guess position. The tracker sequentially attempts search boxes of $4^\circ$,
$3^\circ$, $2^\circ$, and $1.5^\circ$ radius until a valid interior local minimum is identified. If all MSL-based attempts fail, the procedure falls back to detecting the closest local minimum of Z700, and then repeats the procedure within varying box sizes for MSL.

Following the ECMWF's TC tracker algorithm described by Van Der Grijn \cite{der2002tropical}, FuXi-Extreme tracking method \cite{zhong2024fuxi-exteme}, and informed by our analysis of mistracks within our dataset, we introduce several enhancements to this algorithm as described below.

(1)In addition to utilizing the local minima of MSL and Z700, we incorporate two physical termination criteria based on the local maxima of WS10M and vorticity at 850 hPa to enhance the robustness of TC identification. Specifically, tracking is terminated if the maximum 850 hPa relative vorticity within a $1.5^\circ$ range of the center falls below $5 \times 10^{-5}$ s$^{-1}$ or if the maximum WS10M drops below 8 m/s within the same range. This thresholding prevents spurious tracking artifacts that may arise when the cyclone's intensity dissipates.

(2)The models evaluated in our study operate at varying spatial resolutions, specifically 
\(0.25^\circ\) and \(0.1^\circ\). To account for these differences, we linearly scale the parameters of the Gaussian low-pass filter according to the specific grid spacing of each dataset. This adjustment standardizes the effective smoothing area across all forecasts. Consequently, the physical scale for local minimum detection remains approximately \(2^\circ \times 2\) regardless of the native model resolution, maintaining consistent tracking behavior across forecast systems with different grid spacings.

\subsubsection{TC track and intensity evaluation}
Track accuracy is quantified using the MAE, which calculates distance between observed and predicted TC center positions. Intensity evaluation involves assessing the maximum sustained WS10M near the TC center using MAE. Phases categorized as Tropical Depression (TD) or those flagged as Extratropical Cyclone (ET) stages were explicitly excluded from the dataset to minimize tracking uncertainties associated with weak or asymmetric baroclinic systems. Consistent with the WNP methodology, the evaluation included valid time steps categorized as Tropical Storm, Hurricane, Tropical Cyclone, Typhoon, or Super Typhoon. This evaluation encompasses totally 35 TCs from 2024 in both WNP and NA only if the IBTrACS, European Centre for Medium-Range Weather Forecasts Reanalysis v5 (ERA5) and forecast results are simultaneously recorded.  

\subsubsection{Total precipitation evaluation}
To quantitatively assess the forecast model performance, we utilize standard categorical verification statistics based on contingency table including True Positives (TP), False Positives (FP), False Negatives (FN), and True Negatives (TN) \cite{Wilks2011}. The Critical Success Index (CSI), also known as the Threat Score (TS), provides a comprehensive measure of accuracy, accounting for both misses and false alarms. It is calculated using the equation \(\text{CSI} = \frac{\text{TP}}{\text{TP} + \text{FP} + \text{FN}}\). Probability of Detection (POD) score measures the fraction of observed events that were correctly predicted, using the equation \(\text{POD} = \frac{\text{TP}}{\text{TP} + \text{FN}}\). And False Alarm Ratio (FAR) score highlights the tendency of the dataset to overestimate precipitation occurrence, using the equation \(\text{FAR} = \frac{\text{FP}}{\text{TP} + \text{FP}}\).

The 6-hour accumulated ground truth precipitation data on land were obtained from the AWS provided by the China Meteorological Data Service Centre (CMDC). Since the AWS network is limited to mainland China, a spatial masking procedure was applied to the forecast data. Therefore, grid points corresponding to oceanic regions and areas outside mainland Chinese territory were excluded to ensure spatial consistency.

Furthermore, direct comparison between point-based gauge observations and grid-based NWP outputs is prone to representativeness error due to scale mismatches \cite{Tustison2001}. To mitigate this and facilitate a rigorous grid-to-grid verification, we reference the standard practice to map observational data onto the model grid \cite{accadia2003sensitivity,accadia2005verification}. Therefore, consistent with established verification frameworks, the valid station observations were interpolated to the \(0.1^\circ\) model's grid resolution prior to statistical calculation. The metrics were computed for precipitation accumulated over a 6-hour period, using threshold values of 0.1, 10, 20, and 40 mm to evaluate the model's performance across light to heavy rainfall intensities.

\section*{Data Availability Statement}

The ERA5 reanalysis data are accessible through the Copernicus Climate Data Store at \url{https://cds.climate.copernicus.eu/}.
ECMWF HRES and ENS TC track forecasts can be retrieved from \url{https://confluence.ecmwf.int/display/TIGGE}.
We downloaded the ground truth tracks of TC from the International Best Track Archive for Climate Stewardship (IBTrACS) project, which is publicly available at \url{https://www.ncei.noaa.gov/products/international-best-track-archive}.

\section*{Code Availability Statement}
The code base of the DDPM model used in this work is available at \url{https://github.com/luchengthu/dpm-solver} \cite{lu2022dpmsolverpp}.
The source code used for running FuXi-TC model, along with the scripts for running the WRF simulations in this work, are available from the Zenodo link (\url{https://zenodo.org/records/19382719} \cite{guo_2026_19382719}). The source code used for training and running FuXi models are available at \url{https://doi.org/10.5281/zenodo.8100201} \cite{chen2023fuxi}.
Additionally, the basic Aurora tracker algorithm in this work can be downloaded from \url{https://github.com/microsoft/aurora} \cite{bodnar2025foundation}.
Furthermore, the source code for the WRF model version 4.3 is publicly available at \url{https://github.com/wrf-model/WRF} \cite{jensen2021description}.

\section*{Acknowledgments}
We gratefully acknowledge the ECMWF for their efforts in producing, maintaining, and distributing the ERA5 reanalysis and ECMWF HRES datasets.
We acknowledged the efforts of NOAA National Centers for Environmental Information in making the IBTrACS dataset available.
We recognize the efforts of CMA for sharing the IBTrACS dataset.
We thank support from the Computing for the Future at Fudan (CFFF), which has provided us with a high-performance computing platform.
This work was supported by AI for Science Program, Shanghai Municipal Commission of Economy and Informatization, 2025-GZL-RGZN-BTBX-02017.

\section*{Competing interests}
The authors declare no competing interests.

\noindent

\bibliography{refs}


\begin{thebibliography}{69}
\ifx \bisbn   \undefined \def \bisbn  #1{ISBN #1}\fi
\ifx \binits  \undefined \def \binits#1{#1}\fi
\ifx \bauthor  \undefined \def \bauthor#1{#1}\fi
\ifx \batitle  \undefined \def \batitle#1{#1}\fi
\ifx \bjtitle  \undefined \def \bjtitle#1{#1}\fi
\ifx \bvolume  \undefined \def \bvolume#1{\textbf{#1}}\fi
\ifx \byear  \undefined \def \byear#1{#1}\fi
\ifx \bissue  \undefined \def \bissue#1{#1}\fi
\ifx \bfpage  \undefined \def \bfpage#1{#1}\fi
\ifx \blpage  \undefined \def \blpage #1{#1}\fi
\ifx \burl  \undefined \def \burl#1{\textsf{#1}}\fi
\ifx \doiurl  \undefined \def \doiurl#1{\url{https://doi.org/#1}}\fi
\ifx \betal  \undefined \def \betal{\textit{et al.}}\fi
\ifx \binstitute  \undefined \def \binstitute#1{#1}\fi
\ifx \binstitutionaled  \undefined \def \binstitutionaled#1{#1}\fi
\ifx \bctitle  \undefined \def \bctitle#1{#1}\fi
\ifx \beditor  \undefined \def \beditor#1{#1}\fi
\ifx \bpublisher  \undefined \def \bpublisher#1{#1}\fi
\ifx \bbtitle  \undefined \def \bbtitle#1{#1}\fi
\ifx \bedition  \undefined \def \bedition#1{#1}\fi
\ifx \bseriesno  \undefined \def \bseriesno#1{#1}\fi
\ifx \blocation  \undefined \def \blocation#1{#1}\fi
\ifx \bsertitle  \undefined \def \bsertitle#1{#1}\fi
\ifx \bsnm \undefined \def \bsnm#1{#1}\fi
\ifx \bsuffix \undefined \def \bsuffix#1{#1}\fi
\ifx \bparticle \undefined \def \bparticle#1{#1}\fi
\ifx \barticle \undefined \def \barticle#1{#1}\fi
\bibcommenthead
\ifx \bconfdate \undefined \def \bconfdate #1{#1}\fi
\ifx \botherref \undefined \def \botherref #1{#1}\fi
\ifx \url \undefined \def \url#1{\textsf{#1}}\fi
\ifx \bchapter \undefined \def \bchapter#1{#1}\fi
\ifx \bbook \undefined \def \bbook#1{#1}\fi
\ifx \bcomment \undefined \def \bcomment#1{#1}\fi
\ifx \oauthor \undefined \def \oauthor#1{#1}\fi
\ifx \citeauthoryear \undefined \def \citeauthoryear#1{#1}\fi
\ifx \endbibitem  \undefined \def \endbibitem {}\fi
\ifx \bconflocation  \undefined \def \bconflocation#1{#1}\fi
\ifx \arxivurl  \undefined \def \arxivurl#1{\textsf{#1}}\fi
\csname PreBibitemsHook\endcsname

\bibitem[\protect\citeauthoryear{Emanuel}{2005}]{emanuel2005increasing}
\begin{barticle}
\bauthor{\bsnm{Emanuel}, \binits{K.}}:
\batitle{Increasing destructiveness of tropical cyclones over the past 30 years}.
\bjtitle{Nature}
\bvolume{436}(\bissue{7051}),
\bfpage{686}--\blpage{688}
(\byear{2005})
\doiurl{10.1038/nature03906}
\end{barticle}
\endbibitem

\bibitem[\protect\citeauthoryear{Emanuel et~al.}{2012}]{emanuel2012impact}
\begin{botherref}
\oauthor{\bsnm{Emanuel}, \binits{K.}},
\oauthor{\bsnm{Chonabayashi}, \binits{S.}},
\oauthor{\bsnm{Bakkensen}, \binits{L.}},
\oauthor{\bsnm{Mendelsohn}, \binits{R.}}:
The impact of climate change on global tropical cyclone damage.
Nature Climate Change
\textbf{2}
(2012)
\doiurl{10.1038/nclimate1357}
\end{botherref}
\endbibitem

\bibitem[\protect\citeauthoryear{Zhang et~al.}{2009}]{zhang2009tropical}
\begin{barticle}
\bauthor{\bsnm{Zhang}, \binits{Q.}},
\bauthor{\bsnm{Wu}, \binits{L.}},
\bauthor{\bsnm{Liu}, \binits{Q.}}:
\batitle{Tropical cyclone damages in china 1983--2006}.
\bjtitle{Bulletin of the American Meteorological Society}
\bvolume{90}(\bissue{4}),
\bfpage{489}--\blpage{496}
(\byear{2009})
\doiurl{10.1175/2008BAMS2631.1}
\end{barticle}
\endbibitem

\bibitem[\protect\citeauthoryear{Zhang et~al.}{2017}]{zhang2017impact}
\begin{barticle}
\bauthor{\bsnm{Zhang}, \binits{Q.}},
\bauthor{\bsnm{Gu}, \binits{X.}},
\bauthor{\bsnm{Shi}, \binits{P.}},
\bauthor{\bsnm{Singh}, \binits{V.P.}}:
\batitle{Impact of tropical cyclones on flood risk in southeastern china: Spatial patterns, causes and implications}.
\bjtitle{Global and Planetary Change}
\bvolume{150},
\bfpage{81}--\blpage{93}
(\byear{2017})
\doiurl{10.1016/j.gloplacha.2017.02.004}
\end{barticle}
\endbibitem

\bibitem[\protect\citeauthoryear{Peduzzi et~al.}{2012}]{peduzzi2012global}
\begin{barticle}
\bauthor{\bsnm{Peduzzi}, \binits{P.}},
\bauthor{\bsnm{Chatenoux}, \binits{B.}},
\bauthor{\bsnm{Dao}, \binits{H.}},
\bauthor{\bsnm{De~Bono}, \binits{A.}},
\bauthor{\bsnm{Herold}, \binits{C.}},
\bauthor{\bsnm{Kossin}, \binits{J.}},
\bauthor{\bsnm{Mouton}, \binits{F.}},
\bauthor{\bsnm{Nordbeck}, \binits{O.}}:
\batitle{Global trends in tropical cyclone risk}.
\bjtitle{Nature Climate Change}
\bvolume{2}(\bissue{4}),
\bfpage{289}--\blpage{294}
(\byear{2012})
\doiurl{10.1038/nclimate1410}
\end{barticle}
\endbibitem

\bibitem[\protect\citeauthoryear{Masson-Delmotte et~al.}{2021}]{IPCC_AR6_WGI_2021}
\begin{bbook}
\beditor{\bsnm{Masson-Delmotte}, \binits{V.}},
\beditor{\bsnm{Zhai}, \binits{P.}},
\beditor{\bsnm{Pirani}, \binits{A.}},
\beditor{\bsnm{Connors}, \binits{S.L.}},
\beditor{\bsnm{Péan}, \binits{C.}},
\beditor{\bsnm{Berger}, \binits{S.}},
\beditor{\bsnm{Caud}, \binits{N.}},
\beditor{\bsnm{Chen}, \binits{Y.}},
\beditor{\bsnm{Goldfarb}, \binits{L.}},
\beditor{\bsnm{Gomis}, \binits{M.I.}},
\beditor{\bsnm{Huang}, \binits{M.}},
\beditor{\bsnm{Leitzell}, \binits{K.}},
\beditor{\bsnm{Lonnoy}, \binits{E.}},
\beditor{\bsnm{Matthews}, \binits{J.B.R.}},
\beditor{\bsnm{Maycock}, \binits{T.K.}},
\beditor{\bsnm{Waterfield}, \binits{T.}},
\beditor{\bsnm{Yelekçi}, \binits{O.}},
\beditor{\bsnm{Yu}, \binits{R.}},
\beditor{\bsnm{Zhou}, \binits{B.}} (eds.):
\bbtitle{Climate Change 2021: The Physical Science Basis. Contribution of Working Group I to the Sixth Assessment Report of the Intergovernmental Panel on Climate Change}.
\bpublisher{Cambridge University Press}, \blocation{???}
(\byear{2021}).
\doiurl{10.1017/9781009157896} .
\burl{https://www.ipcc.ch/report/ar6/wg1/}
\end{bbook}
\endbibitem

\bibitem[\protect\citeauthoryear{Webster et~al.}{2005}]{webster2005changes}
\begin{barticle}
\bauthor{\bsnm{Webster}, \binits{P.}},
\bauthor{\bsnm{Holland}, \binits{G.}},
\bauthor{\bsnm{Curry}, \binits{J.}},
\bauthor{\bsnm{Chang}, \binits{H.-R.}}:
\batitle{Changes in tropical cyclone number, duration, and intensity in a warming environment}.
\bjtitle{Science}
\bvolume{309},
\bfpage{1844}--\blpage{1846}
(\byear{2005})
\doiurl{10.1126/science.1116448}
\end{barticle}
\endbibitem

\bibitem[\protect\citeauthoryear{Tallapragada}{2016}]{tallapragada2016hwrf}
\begin{bbook}
\bauthor{\bsnm{Tallapragada}, \binits{V.}}:
In: \beditor{\bsnm{Mohanty}, \binits{U.C.}},
\beditor{\bsnm{Gopalakrishnan}, \binits{S.G.}} (eds.)
\bbtitle{Overview of the NOAA/NCEP Operational Hurricane Weather Research and Forecast (HWRF) Modelling System},
pp. \bfpage{51}--\blpage{106}.
\bpublisher{Springer}, \blocation{???}
(\byear{2016}).
\doiurl{10.5822/978-94-024-0896-6\_3} .
\burl{https://doi.org/10.5822/978-94-024-0896-6\_3}
\end{bbook}
\endbibitem

\bibitem[\protect\citeauthoryear{Alaka et~al.}{2024}]{alaka2024hwrf}
\begin{barticle}
\bauthor{\bsnm{Alaka}, \binits{G.J.}},
\bauthor{\bsnm{Sippel}, \binits{J.A.}},
\bauthor{\bsnm{Zhang}, \binits{Z.}},
\bauthor{\bsnm{Kim}, \binits{H.-S.}},
\bauthor{\bsnm{Marks}, \binits{F.D.}},
\bauthor{\bsnm{Tallapragada}, \binits{V.}},
\bauthor{\bsnm{Mehra}, \binits{A.}},
\bauthor{\bsnm{Zhang}, \binits{X.}},
\bauthor{\bsnm{Poyer}, \binits{A.}},
\bauthor{\bsnm{Gopalakrishnan}, \binits{S.G.}}:
\batitle{Lifetime performance of the operational hurricane weather research and forecasting model (hwrf) for north atlantic tropical cyclones}.
\bjtitle{Bulletin of the American Meteorological Society}
\bvolume{105}(\bissue{6}),
\bfpage{932}--\blpage{961}
(\byear{2024})
\doiurl{10.1175/BAMS-D-23-0139.1}
\end{barticle}
\endbibitem

\bibitem[\protect\citeauthoryear{Wang et~al.}{2023}]{wang2023hafs}
\begin{barticle}
\bauthor{\bsnm{Wang}, \binits{W.}},
\bauthor{\bsnm{Han}, \binits{J.}},
\bauthor{\bsnm{Yang}, \binits{F.}},
\bauthor{\bsnm{Steffen}, \binits{J.}},
\bauthor{\bsnm{Liu}, \binits{B.}},
\bauthor{\bsnm{Zhang}, \binits{Z.}},
\bauthor{\bsnm{Mehra}, \binits{A.}},
\bauthor{\bsnm{Tallapragada}, \binits{V.}}:
\batitle{Improving the intensity forecast of tropical cyclones in the hurricane analysis and forecast system}.
\bjtitle{Weather and Forecasting}
\bvolume{38}(\bissue{10}),
\bfpage{2057}--\blpage{2075}
(\byear{2023})
\doiurl{10.1175/WAF-D-23-0041.1}
\end{barticle}
\endbibitem

\bibitem[\protect\citeauthoryear{Ma et~al.}{2021}]{ma2021grapes}
\begin{barticle}
\bauthor{\bsnm{Ma}, \binits{S.}},
\bauthor{\bsnm{Zhang}, \binits{J.}},
\bauthor{\bsnm{Qu}, \binits{A.}},
\bauthor{\bsnm{Wang}, \binits{D.}},
\bauthor{\bsnm{Shen}, \binits{X.}}:
\batitle{Impacts to tropical cyclone prediction of grapes\_tym from increasing of model vertical levels and enlargement of model forecast domain}.
\bjtitle{Acta Meteorologica Sinica}
\bvolume{79}(\bissue{1}),
\bfpage{94}--\blpage{103}
(\byear{2021})
\doiurl{10.11676/qxxb2020.067}
\end{barticle}
\endbibitem

\bibitem[\protect\citeauthoryear{Cangialosi et~al.}{2020}]{cangialosi2020progress}
\begin{barticle}
\bauthor{\bsnm{Cangialosi}, \binits{J.P.}},
\bauthor{\bsnm{Blake}, \binits{E.}},
\bauthor{\bsnm{DeMaria}, \binits{M.}},
\bauthor{\bsnm{Penny}, \binits{A.}},
\bauthor{\bsnm{Latto}, \binits{A.}},
\bauthor{\bsnm{Rappaport}, \binits{E.}},
\bauthor{\bsnm{Tallapragada}, \binits{V.}}:
\batitle{Recent progress in tropical cyclone intensity forecasting at the national hurricane center}.
\bjtitle{Weather and Forecasting}
\bvolume{35}(\bissue{5}),
\bfpage{1913}--\blpage{1922}
(\byear{2020})
\doiurl{10.1175/WAF-D-20-0059.1}
\end{barticle}
\endbibitem

\bibitem[\protect\citeauthoryear{Xu et~al.}{2025}]{xu2025exploring}
\begin{barticle}
\bauthor{\bsnm{Xu}, \binits{H.}},
\bauthor{\bsnm{Zhao}, \binits{Y.}},
\bauthor{\bsnm{Zhao}, \binits{D.}},
\bauthor{\bsnm{Duan}, \binits{Y.}},
\bauthor{\bsnm{Xu}, \binits{X.}}:
\batitle{Exploring the typhoon intensity forecasting through integrating ai weather forecasting with regional numerical weather model}.
\bjtitle{npj Climate and Atmospheric Science}
\bvolume{8}(\bissue{1}),
\bfpage{38}
(\byear{2025})
\end{barticle}
\endbibitem

\bibitem[\protect\citeauthoryear{Xinyuan et~al.}{2025}]{xinyuan2025review}
\begin{botherref}
\oauthor{\bsnm{Xinyuan}, \binits{B.}},
\oauthor{\bsnm{Jinping}, \binits{L.}},
\oauthor{\bsnm{Yihong}, \binits{D.}}:
Review of artificial intelligence application in typhoon forecasting.
Tropical Cyclone Research and Review
(2025)
\end{botherref}
\endbibitem

\bibitem[\protect\citeauthoryear{Bi et~al.}{2023}]{bi2023accurate}
\begin{barticle}
\bauthor{\bsnm{Bi}, \binits{K.}},
\bauthor{\bsnm{Xie}, \binits{L.}},
\bauthor{\bsnm{Zhang}, \binits{H.}},
\bauthor{\bsnm{Chen}, \binits{X.}},
\bauthor{\bsnm{Gu}, \binits{X.}},
\bauthor{\bsnm{Tian}, \binits{Q.}}:
\batitle{Accurate medium-range global weather forecasting with 3d neural networks}.
\bjtitle{Nature}
\bvolume{619}(\bissue{7970}),
\bfpage{533}--\blpage{538}
(\byear{2023})
\end{barticle}
\endbibitem

\bibitem[\protect\citeauthoryear{Zhong et~al.}{2024}]{zhong2024fuxi-exteme}
\begin{botherref}
\oauthor{\bsnm{Zhong}, \binits{X.}},
\oauthor{\bsnm{Chen}, \binits{L.}},
\oauthor{\bsnm{Liu}, \binits{J.}},
\oauthor{\bsnm{Lin}, \binits{C.}},
\oauthor{\bsnm{Qi}, \binits{Y.}},
\oauthor{\bsnm{Li}, \binits{H.}}:
Fuxi-extreme: Improving extreme rainfall and wind forecasts with diffusion model.
Science China Earth Sciences,
1--13
(2024)
\end{botherref}
\endbibitem

\bibitem[\protect\citeauthoryear{Mahesh et~al.}{2023}]{mahesh2023evaluating}
\begin{bchapter}
\bauthor{\bsnm{Mahesh}, \binits{A.}},
\bauthor{\bsnm{Cohen}, \binits{Y.}},
\bauthor{\bsnm{Brenowitz}, \binits{N.}},
\bauthor{\bsnm{Elms}, \binits{J.}},
\bauthor{\bsnm{Subramanian}, \binits{S.}},
\bauthor{\bsnm{Harrington}, \binits{P.}},
\bauthor{\bsnm{Anandkumar}, \binits{A.}},
\bauthor{\bsnm{Pathak}, \binits{J.}},
\bauthor{\bsnm{Kurth}, \binits{T.}},
\bauthor{\bsnm{Bonev}, \binits{B.}}, \betal:
\bctitle{Evaluating data-driven forecasts of extreme weather}.
In: \bbtitle{AGU Fall Meeting Abstracts},
vol. \bseriesno{2023},
pp. \bfpage{31}--\blpage{2513}
(\byear{2023})
\end{bchapter}
\endbibitem

\bibitem[\protect\citeauthoryear{Wang et~al.}{2024}]{wang2024machine}
\begin{bchapter}
\bauthor{\bsnm{Wang}, \binits{J.}},
\bauthor{\bsnm{Tabas}, \binits{S.}},
\bauthor{\bsnm{Yang}, \binits{F.}},
\bauthor{\bsnm{Levit}, \binits{J.}},
\bauthor{\bsnm{Stajner}, \binits{I.}},
\bauthor{\bsnm{Montuoro}, \binits{R.}},
\bauthor{\bsnm{Tallapragada}, \binits{V.}},
\bauthor{\bsnm{Gross}, \binits{B.}}:
\bctitle{Machine learning weather prediction model development for global ensemble forecasts at ncep}.
In: \bbtitle{EGU General Assembly Conference Abstracts},
p. \bfpage{11707}
(\byear{2024})
\end{bchapter}
\endbibitem

\bibitem[\protect\citeauthoryear{Shi et~al.}{2025}]{shi2025comparison}
\begin{barticle}
\bauthor{\bsnm{Shi}, \binits{Y.}},
\bauthor{\bsnm{Hu}, \binits{R.}},
\bauthor{\bsnm{Wu}, \binits{N.}},
\bauthor{\bsnm{Zhang}, \binits{H.}},
\bauthor{\bsnm{Liu}, \binits{X.}},
\bauthor{\bsnm{Zeng}, \binits{Z.}},
\bauthor{\bsnm{Zhu}, \binits{J.}},
\bauthor{\bsnm{Han}, \binits{P.}},
\bauthor{\bsnm{Luo}, \binits{C.}},
\bauthor{\bsnm{Zhang}, \binits{H.}}, \betal:
\batitle{Comparison of ai and nwp models in operational severe weather forecasting: A study on tropical cyclone predictions}.
\bjtitle{Journal of Geophysical Research: Machine Learning and Computation}
\bvolume{2}(\bissue{2}),
\bfpage{2024}--\blpage{000481}
(\byear{2025})
\end{barticle}
\endbibitem

\bibitem[\protect\citeauthoryear{Liu et~al.}{2024}]{liu2024hybrid}
\begin{barticle}
\bauthor{\bsnm{Liu}, \binits{H.-Y.}},
\bauthor{\bsnm{Tan}, \binits{Z.-M.}},
\bauthor{\bsnm{Wang}, \binits{Y.}},
\bauthor{\bsnm{Tang}, \binits{J.}},
\bauthor{\bsnm{Satoh}, \binits{M.}},
\bauthor{\bsnm{Lei}, \binits{L.}},
\bauthor{\bsnm{Gu}, \binits{J.-F.}},
\bauthor{\bsnm{Zhang}, \binits{Y.}},
\bauthor{\bsnm{Nie}, \binits{G.-Z.}},
\bauthor{\bsnm{Chen}, \binits{Q.-Z.}}:
\batitle{A hybrid machine learning/physics-based modeling framework for 2-week extended prediction of tropical cyclones}.
\bjtitle{Journal of Geophysical Research: Machine Learning and Computation}
\bvolume{1}(\bissue{3}),
\bfpage{2024}--\blpage{000207}
(\byear{2024})
\end{barticle}
\endbibitem

\bibitem[\protect\citeauthoryear{Hersbach et~al.}{2020}]{hersbach2020era5}
\begin{barticle}
\bauthor{\bsnm{Hersbach}, \binits{H.}},
\bauthor{\bsnm{Bell}, \binits{B.}},
\bauthor{\bsnm{Berrisford}, \binits{P.}},
\bauthor{\bsnm{Hirahara}, \binits{S.}},
\bauthor{\bsnm{Hor{\'a}nyi}, \binits{A.}},
\bauthor{\bsnm{Mu{\~n}oz-Sabater}, \binits{J.}},
\bauthor{\bsnm{Nicolas}, \binits{J.}},
\bauthor{\bsnm{Peubey}, \binits{C.}},
\bauthor{\bsnm{Radu}, \binits{R.}},
\bauthor{\bsnm{Schepers}, \binits{D.}}, \betal:
\batitle{The era5 global reanalysis}.
\bjtitle{Quarterly journal of the royal meteorological society}
\bvolume{146}(\bissue{730}),
\bfpage{1999}--\blpage{2049}
(\byear{2020})
\end{barticle}
\endbibitem

\bibitem[\protect\citeauthoryear{Hodges et~al.}{2017}]{Hodges2017}
\begin{barticle}
\bauthor{\bsnm{Hodges}, \binits{K.}},
\bauthor{\bsnm{Cobb}, \binits{A.}},
\bauthor{\bsnm{Vidale}, \binits{P.L.}}:
\batitle{How well are tropical cyclones represented in reanalysis datasets?}
\bjtitle{Journal of Climate}
\bvolume{30}(\bissue{14}),
\bfpage{5243}--\blpage{5264}
(\byear{2017})
\doiurl{10.1175/JCLI-D-16-0557.1}
\end{barticle}
\endbibitem

\bibitem[\protect\citeauthoryear{Zhang et~al.}{2024}]{zhang2024typhoon}
\begin{barticle}
\bauthor{\bsnm{Zhang}, \binits{X.}},
\bauthor{\bsnm{Zuo}, \binits{C.}},
\bauthor{\bsnm{Wang}, \binits{Z.}},
\bauthor{\bsnm{Tao}, \binits{C.}},
\bauthor{\bsnm{Han}, \binits{Y.}},
\bauthor{\bsnm{Zuo}, \binits{J.}}:
\batitle{Typhoon storm surge simulation study based on reconstructed era5 wind fields—a case study of typhoon “muifa”, the 12th typhoon of 2022}.
\bjtitle{Journal of Marine Science and Engineering}
\bvolume{12}(\bissue{11}),
\bfpage{2099}
(\byear{2024})
\end{barticle}
\endbibitem

\bibitem[\protect\citeauthoryear{Knapp et~al.}{2010}]{knapp2010international}
\begin{barticle}
\bauthor{\bsnm{Knapp}, \binits{K.R.}},
\bauthor{\bsnm{Kruk}, \binits{M.C.}},
\bauthor{\bsnm{Levinson}, \binits{D.H.}},
\bauthor{\bsnm{Diamond}, \binits{H.J.}},
\bauthor{\bsnm{Neumann}, \binits{C.J.}}:
\batitle{The international best track archive for climate stewardship (ibtracs) unifying tropical cyclone data}.
\bjtitle{Bulletin of the American Meteorological Society}
\bvolume{91}(\bissue{3}),
\bfpage{363}--\blpage{376}
(\byear{2010})
\end{barticle}
\endbibitem

\bibitem[\protect\citeauthoryear{Wang et~al.}{2025}]{wang2025vqlti}
\begin{botherref}
\oauthor{\bsnm{Wang}, \binits{X.}},
\oauthor{\bsnm{Liu}, \binits{L.}},
\oauthor{\bsnm{Chen}, \binits{K.}},
\oauthor{\bsnm{Han}, \binits{T.}},
\oauthor{\bsnm{Li}, \binits{B.}},
\oauthor{\bsnm{Bai}, \binits{L.}}:
Vqlti: Long-term tropical cyclone intensity forecasting with physical constraints.
arXiv preprint arXiv:2501.18122
(2025)
\end{botherref}
\endbibitem

\bibitem[\protect\citeauthoryear{Huang et~al.}{2025}]{huang2025benchmark}
\begin{barticle}
\bauthor{\bsnm{Huang}, \binits{C.}},
\bauthor{\bsnm{Mu}, \binits{P.}},
\bauthor{\bsnm{Zhang}, \binits{J.}},
\bauthor{\bsnm{Chan}, \binits{S.}},
\bauthor{\bsnm{Zhang}, \binits{S.}},
\bauthor{\bsnm{Yan}, \binits{H.}},
\bauthor{\bsnm{Chen}, \binits{S.}},
\bauthor{\bsnm{Bai}, \binits{C.}}:
\batitle{Benchmark dataset and deep learning method for global tropical cyclone forecasting}.
\bjtitle{Nature Communications}
\bvolume{16}(\bissue{1}),
\bfpage{5923}
(\byear{2025})
\end{barticle}
\endbibitem

\bibitem[\protect\citeauthoryear{Husain et~al.}{2025}]{husain2025leveraging}
\begin{botherref}
\oauthor{\bsnm{Husain}, \binits{S.Z.}},
\oauthor{\bsnm{Separovic}, \binits{L.}},
\oauthor{\bsnm{Caron}, \binits{J.-F.}},
\oauthor{\bsnm{Aider}, \binits{R.}},
\oauthor{\bsnm{Buehner}, \binits{M.}},
\oauthor{\bsnm{Chamberland}, \binits{S.}},
\oauthor{\bsnm{Lapalme}, \binits{E.}},
\oauthor{\bsnm{McTaggart-Cowan}, \binits{R.}},
\oauthor{\bsnm{Subich}, \binits{C.}},
\oauthor{\bsnm{Vaillancourt}, \binits{P.A.}}, et al.:
Leveraging data-driven weather models for improving numerical weather prediction skill through large-scale spectral nudging.
Weather and Forecasting
\textbf{1}(aop)
(2025)
\end{botherref}
\endbibitem

\bibitem[\protect\citeauthoryear{Sohl-Dickstein et~al.}{2015}]{sohl2015deep}
\begin{bchapter}
\bauthor{\bsnm{Sohl-Dickstein}, \binits{J.}},
\bauthor{\bsnm{Weiss}, \binits{E.}},
\bauthor{\bsnm{Maheswaranathan}, \binits{N.}},
\bauthor{\bsnm{Ganguli}, \binits{S.}}:
\bctitle{Deep unsupervised learning using nonequilibrium thermodynamics}.
In: \bbtitle{International Conference on Machine Learning},
pp. \bfpage{2256}--\blpage{2265}
(\byear{2015}).
\bcomment{PMLR}
\end{bchapter}
\endbibitem

\bibitem[\protect\citeauthoryear{Ho et~al.}{2020}]{ho2020denoising}
\begin{bchapter}
\bauthor{\bsnm{Ho}, \binits{J.}},
\bauthor{\bsnm{Jain}, \binits{A.}},
\bauthor{\bsnm{Abbeel}, \binits{P.}}:
\bctitle{Denoising diffusion probabilistic models}.
In: \bbtitle{Advances in Neural Information Processing Systems},
vol. \bseriesno{33},
pp. \bfpage{6840}--\blpage{6851}
(\byear{2020})
\end{bchapter}
\endbibitem

\bibitem[\protect\citeauthoryear{Chen et~al.}{2023}]{chen2023fuxi}
\begin{barticle}
\bauthor{\bsnm{Chen}, \binits{L.}},
\bauthor{\bsnm{Zhong}, \binits{X.}},
\bauthor{\bsnm{Zhang}, \binits{F.}},
\bauthor{\bsnm{Cheng}, \binits{Y.}},
\bauthor{\bsnm{Xu}, \binits{Y.}},
\bauthor{\bsnm{Qi}, \binits{Y.}},
\bauthor{\bsnm{Li}, \binits{H.}}:
\batitle{Fuxi: a cascade machine learning forecasting system for 15-day global weather forecast}.
\bjtitle{npj climate and atmospheric science}
\bvolume{6}(\bissue{1}),
\bfpage{190}
(\byear{2023})
\end{barticle}
\endbibitem

\bibitem[\protect\citeauthoryear{Zhong et~al.}{2024}]{zhong2024fuxiens}
\begin{botherref}
\oauthor{\bsnm{Zhong}, \binits{X.}},
\oauthor{\bsnm{Chen}, \binits{L.}},
\oauthor{\bsnm{Li}, \binits{H.}},
\oauthor{\bsnm{Liu}, \binits{J.}},
\oauthor{\bsnm{Fan}, \binits{X.}},
\oauthor{\bsnm{Feng}, \binits{J.}},
\oauthor{\bsnm{Dai}, \binits{K.}},
\oauthor{\bsnm{Luo}, \binits{J.-J.}},
\oauthor{\bsnm{Wu}, \binits{J.}},
\oauthor{\bsnm{Lu}, \binits{B.}}:
Fuxi-ens: A machine learning model for medium-range ensemble weather forecasting.
arXiv preprint arXiv:2405.05925
(2024)
\end{botherref}
\endbibitem

\bibitem[\protect\citeauthoryear{Wu et~al.}{2025}]{wu2025extreme}
\begin{botherref}
\oauthor{\bsnm{Wu}, \binits{L.}},
\oauthor{\bsnm{Yu}, \binits{R.}},
\oauthor{\bsnm{Xiang}, \binits{C.}},
\oauthor{\bsnm{Yu}, \binits{H.}},
\oauthor{\bsnm{Feng}, \binits{Y.}},
\oauthor{\bsnm{Zhou}, \binits{X.}}:
Extreme Impacts of Four Landfalling Tropical Cyclones in China during the 2024 Peak Season.
Springer
(2025)
\end{botherref}
\endbibitem

\bibitem[\protect\citeauthoryear{Polichtchouk et~al.}{2026}]{polichtchouk2026hybridensembleforecastingcombining}
\begin{botherref}
\oauthor{\bsnm{Polichtchouk}, \binits{I.}},
\oauthor{\bsnm{Lang}, \binits{S.}},
\oauthor{\bsnm{Lock}, \binits{S.-J.}},
\oauthor{\bsnm{Maier-Gerber}, \binits{M.}},
\oauthor{\bsnm{Dueben}, \binits{P.}}:
Hybrid ensemble forecasting combining physics-based and machine-learning predictions through spectral nudging
(2026).
\url{https://arxiv.org/abs/2603.05570}
\end{botherref}
\endbibitem

\bibitem[\protect\citeauthoryear{Villalobos et~al.}{2024}]{villalobos2024position}
\begin{bchapter}
\bauthor{\bsnm{Villalobos}, \binits{P.}},
\bauthor{\bsnm{Ho}, \binits{A.}},
\bauthor{\bsnm{Sevilla}, \binits{J.}},
\bauthor{\bsnm{Besiroglu}, \binits{T.}},
\bauthor{\bsnm{Heim}, \binits{L.}},
\bauthor{\bsnm{Hobbhahn}, \binits{M.}}:
\bctitle{Position: Will we run out of data? limits of {LLM} scaling based on human-generated data}.
In: \bbtitle{Forty-first International Conference on Machine Learning}
(\byear{2024}).
\burl{https://openreview.net/forum?id=ViZcgDQjyG}
\end{bchapter}
\endbibitem

\bibitem[\protect\citeauthoryear{Shumailov et~al.}{2024}]{shumailov2024ai}
\begin{barticle}
\bauthor{\bsnm{Shumailov}, \binits{I.}},
\bauthor{\bsnm{Shumaylov}, \binits{Z.}},
\bauthor{\bsnm{Zhao}, \binits{Y.}},
\bauthor{\bsnm{Papernot}, \binits{N.}},
\bauthor{\bsnm{Anderson}, \binits{R.}},
\bauthor{\bsnm{Gal}, \binits{Y.}}:
\batitle{Ai models collapse when trained on recursively generated data}.
\bjtitle{Nature}
\bvolume{631}(\bissue{8022}),
\bfpage{755}--\blpage{759}
(\byear{2024})
\doiurl{10.1038/s41586-024-07566-y}
\end{barticle}
\endbibitem

\bibitem[\protect\citeauthoryear{Gerstgrasser et~al.}{2024}]{gerstgrasser2024model}
\begin{botherref}
\oauthor{\bsnm{Gerstgrasser}, \binits{M.}},
\oauthor{\bsnm{Schaeffer}, \binits{R.}},
\oauthor{\bsnm{Dey}, \binits{A.}},
\oauthor{\bsnm{Rafailov}, \binits{R.}},
\oauthor{\bsnm{Sleight}, \binits{H.}},
\oauthor{\bsnm{Hughes}, \binits{J.}},
\oauthor{\bsnm{Korbak}, \binits{T.}},
\oauthor{\bsnm{Agrawal}, \binits{R.}},
\oauthor{\bsnm{Pai}, \binits{D.}},
\oauthor{\bsnm{Gromov}, \binits{A.}}, et al.:
Is model collapse inevitable? breaking the curse of recursion by accumulating real and synthetic data.
Preprint at https://arxiv.org/abs/2404.01413
(2024)
\end{botherref}
\endbibitem

\bibitem[\protect\citeauthoryear{Bauer}{2024}]{bauer2024if}
\begin{barticle}
\bauthor{\bsnm{Bauer}, \binits{P.}}:
\batitle{What if? numerical weather prediction at the crossroads}.
\bjtitle{Journal of the European Meteorological Society}
\bvolume{1},
\bfpage{100002}
(\byear{2024})
\end{barticle}
\endbibitem

\bibitem[\protect\citeauthoryear{Kenneth et~al.}{2019}]{kenneth2019international}
\begin{botherref}
\oauthor{\bsnm{Kenneth}, \binits{R.}},
\oauthor{\bsnm{Howard}, \binits{J.}},
\oauthor{\bsnm{James}, \binits{P.}},
\oauthor{\bsnm{Michael}, \binits{C.}},
\oauthor{\bsnm{Carl}, \binits{J.}}:
International best track archive for climate stewardship (ibtracs) project, version 4.
(No Title)
(2019)
\end{botherref}
\endbibitem

\bibitem[\protect\citeauthoryear{Bodnar et~al.}{2025}]{bodnar2025foundation}
\begin{barticle}
\bauthor{\bsnm{Bodnar}, \binits{C.}},
\bauthor{\bsnm{Bruinsma}, \binits{W.P.}},
\bauthor{\bsnm{Lucic}, \binits{A.}},
\bauthor{\bsnm{Stanley}, \binits{M.}},
\bauthor{\bsnm{Allen}, \binits{A.}},
\bauthor{\bsnm{Brandstetter}, \binits{J.}},
\bauthor{\bsnm{Garvan}, \binits{P.}},
\bauthor{\bsnm{Riechert}, \binits{M.}},
\bauthor{\bsnm{Weyn}, \binits{J.A.}},
\bauthor{\bsnm{Dong}, \binits{H.}}, \betal:
\batitle{A foundation model for the earth system}.
\bjtitle{Nature}
\bvolume{641}(\bissue{8065}),
\bfpage{1180}--\blpage{1187}
(\byear{2025})
\end{barticle}
\endbibitem

\bibitem[\protect\citeauthoryear{Bougeault et~al.}{2010}]{bougeault2010thorpex}
\begin{barticle}
\bauthor{\bsnm{Bougeault}, \binits{P.}},
\bauthor{\bsnm{Toth}, \binits{Z.}},
\bauthor{\bsnm{Bishop}, \binits{C.}},
\bauthor{\bsnm{Brown}, \binits{B.}},
\bauthor{\bsnm{Burridge}, \binits{D.}},
\bauthor{\bsnm{Chen}, \binits{D.H.}},
\bauthor{\bsnm{Ebert}, \binits{B.}},
\bauthor{\bsnm{Fuentes}, \binits{M.}},
\bauthor{\bsnm{Hamill}, \binits{T.M.}},
\bauthor{\bsnm{Mylne}, \binits{K.}}, \betal:
\batitle{The thorpex interactive grand global ensemble}.
\bjtitle{Bulletin of the American Meteorological Society}
\bvolume{91}(\bissue{8}),
\bfpage{1059}--\blpage{1072}
(\byear{2010})
\end{barticle}
\endbibitem

\bibitem[\protect\citeauthoryear{Swinbank et~al.}{2016}]{swinbank2016tigge}
\begin{barticle}
\bauthor{\bsnm{Swinbank}, \binits{R.}},
\bauthor{\bsnm{Kyouda}, \binits{M.}},
\bauthor{\bsnm{Buchanan}, \binits{P.}},
\bauthor{\bsnm{Froude}, \binits{L.}},
\bauthor{\bsnm{Hamill}, \binits{T.M.}},
\bauthor{\bsnm{Hewson}, \binits{T.D.}},
\bauthor{\bsnm{Keller}, \binits{J.H.}},
\bauthor{\bsnm{Matsueda}, \binits{M.}},
\bauthor{\bsnm{Methven}, \binits{J.}},
\bauthor{\bsnm{Pappenberger}, \binits{F.}}, \betal:
\batitle{The tigge project and its achievements}.
\bjtitle{Bulletin of the American Meteorological Society}
\bvolume{97}(\bissue{1}),
\bfpage{49}--\blpage{67}
(\byear{2016})
\end{barticle}
\endbibitem

\bibitem[\protect\citeauthoryear{Zhong et~al.}{2024}]{zhong2024fuxi2.0}
\begin{botherref}
\oauthor{\bsnm{Zhong}, \binits{X.}},
\oauthor{\bsnm{Chen}, \binits{L.}},
\oauthor{\bsnm{Fan}, \binits{X.}},
\oauthor{\bsnm{Qian}, \binits{W.}},
\oauthor{\bsnm{Liu}, \binits{J.}},
\oauthor{\bsnm{Li}, \binits{H.}}:
Fuxi-2.0: Advancing machine learning weather forecasting model for practical applications.
arXiv preprint arXiv:2409.07188
(2024)
\end{botherref}
\endbibitem

\bibitem[\protect\citeauthoryear{Wallace and Hobbs}{2006}]{wallace2006atmospheric}
\begin{bbook}
\bauthor{\bsnm{Wallace}, \binits{J.M.}},
\bauthor{\bsnm{Hobbs}, \binits{P.V.}}:
\bbtitle{Atmospheric Science: an Introductory Survey},
\bedition{2nd} edn.
\bpublisher{Elsevier Academic Press}, \blocation{???}
(\byear{2006})
\end{bbook}
\endbibitem

\bibitem[\protect\citeauthoryear{Bolton}{1980}]{bolton1980computation}
\begin{barticle}
\bauthor{\bsnm{Bolton}, \binits{D.}}:
\batitle{The computation of equivalent potential temperature}.
\bjtitle{Monthly weather review}
\bvolume{108}(\bissue{7}),
\bfpage{1046}--\blpage{1053}
(\byear{1980})
\end{barticle}
\endbibitem

\bibitem[\protect\citeauthoryear{Niu et~al.}{2025}]{niu2025improving}
\begin{barticle}
\bauthor{\bsnm{Niu}, \binits{Z.}},
\bauthor{\bsnm{Huang}, \binits{W.}},
\bauthor{\bsnm{Zhang}, \binits{L.}},
\bauthor{\bsnm{Deng}, \binits{L.}},
\bauthor{\bsnm{Wang}, \binits{H.}},
\bauthor{\bsnm{Yang}, \binits{Y.}},
\bauthor{\bsnm{Wang}, \binits{D.}},
\bauthor{\bsnm{Li}, \binits{H.}}:
\batitle{Improving typhoon predictions by integrating data-driven machine learning model with physics model based on the spectral nudging and data assimilation}.
\bjtitle{Earth and Space Science}
\bvolume{12}(\bissue{2}),
\bfpage{2024}--\blpage{003952}
(\byear{2025})
\end{barticle}
\endbibitem

\bibitem[\protect\citeauthoryear{Kuhn and Johnson}{2013}]{kuhn2013applied}
\begin{bbook}
\bauthor{\bsnm{Kuhn}, \binits{M.}},
\bauthor{\bsnm{Johnson}, \binits{K.}}:
\bbtitle{Applied Predictive Modeling}
vol. \bseriesno{26}.
\bpublisher{Springer}, \blocation{???}
(\byear{2013})
\end{bbook}
\endbibitem

\bibitem[\protect\citeauthoryear{Shrestha et~al.}{2019}]{shrestha2019evaluation}
\begin{barticle}
\bauthor{\bsnm{Shrestha}, \binits{A.}}, \betal:
\batitle{Evaluation of machine learning models for the routing of ensemble precipitation forecasts}.
\bjtitle{Journal of Hydrology}
\bvolume{570},
\bfpage{646}--\blpage{657}
(\byear{2019})
\end{barticle}
\endbibitem

\bibitem[\protect\citeauthoryear{Saharia et~al.}{2022}]{saharia2022image}
\begin{barticle}
\bauthor{\bsnm{Saharia}, \binits{C.}},
\bauthor{\bsnm{Ho}, \binits{J.}},
\bauthor{\bsnm{Chan}, \binits{W.}},
\bauthor{\bsnm{Salimans}, \binits{T.}},
\bauthor{\bsnm{Fleet}, \binits{D.J.}},
\bauthor{\bsnm{Norouzi}, \binits{M.}}:
\batitle{Image super-resolution via iterative refinement}.
\bjtitle{IEEE Transactions on Pattern Analysis and Machine Intelligence}
\bvolume{45}(\bissue{4}),
\bfpage{4713}--\blpage{4726}
(\byear{2022})
\end{barticle}
\endbibitem

\bibitem[\protect\citeauthoryear{Ronneberger et~al.}{2015}]{Ronneberger2015}
\begin{bchapter}
\bauthor{\bsnm{Ronneberger}, \binits{O.}},
\bauthor{\bsnm{Fischer}, \binits{P.}},
\bauthor{\bsnm{Brox}, \binits{T.}}:
\bctitle{U-net: Convolutional networks for biomedical image segmentation}.
In: \beditor{\bsnm{Navab}, \binits{N.}},
\beditor{\bsnm{Hornegger}, \binits{J.}},
\beditor{\bsnm{Wells}, \binits{W.M.}},
\beditor{\bsnm{Frangi}, \binits{A.F.}} (eds.)
\bbtitle{Medical Image Computing and Computer-Assisted Intervention -- MICCAI 2015},
pp. \bfpage{234}--\blpage{241}.
\bpublisher{Springer},
\blocation{Cham}
(\byear{2015})
\end{bchapter}
\endbibitem

\bibitem[\protect\citeauthoryear{Ramachandran et~al.}{2018}]{ramachandran2017searching}
\begin{bchapter}
\bauthor{\bsnm{Ramachandran}, \binits{P.}},
\bauthor{\bsnm{Zoph}, \binits{B.}},
\bauthor{\bsnm{Le}, \binits{Q.V.}}:
\bctitle{Searching for activation functions}.
In: \bbtitle{International Conference on Learning Representations}
(\byear{2018})
\end{bchapter}
\endbibitem

\bibitem[\protect\citeauthoryear{Wu and He}{2018}]{wu2018group}
\begin{bchapter}
\bauthor{\bsnm{Wu}, \binits{Y.}},
\bauthor{\bsnm{He}, \binits{K.}}:
\bctitle{Group normalization}.
In: \bbtitle{Proceedings of the European Conference on Computer Vision (ECCV)},
pp. \bfpage{3}--\blpage{19}
(\byear{2018})
\end{bchapter}
\endbibitem

\bibitem[\protect\citeauthoryear{Vaswani et~al.}{2017}]{vaswani2017attention}
\begin{bchapter}
\bauthor{\bsnm{Vaswani}, \binits{A.}},
\bauthor{\bsnm{Shazeer}, \binits{N.}},
\bauthor{\bsnm{Parmar}, \binits{N.}},
\bauthor{\bsnm{Uszkoreit}, \binits{J.}},
\bauthor{\bsnm{Jones}, \binits{L.}},
\bauthor{\bsnm{Gomez}, \binits{A.N.}},
\bauthor{\bsnm{Kaiser}, \binits{{\L}.}},
\bauthor{\bsnm{Polosukhin}, \binits{I.}}:
\bctitle{Attention is all you need}.
In: \bbtitle{Advances in Neural Information Processing Systems},
vol. \bseriesno{30}
(\byear{2017})
\end{bchapter}
\endbibitem

\bibitem[\protect\citeauthoryear{Nichol and Dhariwal}{2021}]{nichol2021improved}
\begin{bchapter}
\bauthor{\bsnm{Nichol}, \binits{A.Q.}},
\bauthor{\bsnm{Dhariwal}, \binits{P.}}:
\bctitle{Improved denoising diffusion probabilistic models}.
In: \bbtitle{International Conference on Machine Learning},
pp. \bfpage{8162}--\blpage{8171}
(\byear{2021}).
\bcomment{PMLR}
\end{bchapter}
\endbibitem

\bibitem[\protect\citeauthoryear{Paszke et~al.}{2019}]{paszke2019pytorch}
\begin{bchapter}
\bauthor{\bsnm{Paszke}, \binits{A.}},
\bauthor{\bsnm{Gross}, \binits{S.}},
\bauthor{\bsnm{Massa}, \binits{F.}},
\bauthor{\bsnm{Lerer}, \binits{A.}},
\bauthor{\bsnm{Bradbury}, \binits{J.}},
\bauthor{\bsnm{Chanan}, \binits{G.}},
\bauthor{\bsnm{Killeen}, \binits{T.}},
\bauthor{\bsnm{Lin}, \binits{Z.}},
\bauthor{\bsnm{Gimelshein}, \binits{N.}},
\bauthor{\bsnm{Antiga}, \binits{L.}}, \betal:
\bctitle{Pytorch: An imperative style, high-performance deep learning library}.
In: \bbtitle{Advances in Neural Information Processing Systems},
vol. \bseriesno{32}
(\byear{2019})
\end{bchapter}
\endbibitem

\bibitem[\protect\citeauthoryear{Loshchilov and Hutter}{2019}]{loshchilov2017decoupled}
\begin{bchapter}
\bauthor{\bsnm{Loshchilov}, \binits{I.}},
\bauthor{\bsnm{Hutter}, \binits{F.}}:
\bctitle{Decoupled weight decay regularization}.
In: \bbtitle{International Conference on Learning Representations}
(\byear{2019})
\end{bchapter}
\endbibitem

\bibitem[\protect\citeauthoryear{Jensen et~al.}{2021}]{jensen2021description}
\begin{botherref}
\oauthor{\bsnm{Jensen}, \binits{A.A.}},
\oauthor{\bsnm{Gill}, \binits{D.O.}},
\oauthor{\bsnm{Powers}, \binits{J.G.}},
\oauthor{\bsnm{Duda}, \binits{M.G.}}:
A description of the advanced research wrf model version 4.3.
A Description of the Advanced Research WRF Model Version 4.3 (2021) NCAR/TN-556+ STR
\textbf{556}
(2021)
\end{botherref}
\endbibitem

\bibitem[\protect\citeauthoryear{Niu et~al.}{2025}]{niu2025machine}
\begin{barticle}
\bauthor{\bsnm{Niu}, \binits{Z.}},
\bauthor{\bsnm{Wang}, \binits{D.}},
\bauthor{\bsnm{Mu}, \binits{M.}},
\bauthor{\bsnm{Huang}, \binits{W.}},
\bauthor{\bsnm{Fan}, \binits{X.}},
\bauthor{\bsnm{Yang}, \binits{M.}},
\bauthor{\bsnm{Qin}, \binits{B.}}:
\batitle{Machine-learning (ml)-physics fusion model accelerates the paradigm shift in typhoon forecasting with a cnop-based assimilation framework}.
\bjtitle{Geophysical Research Letters}
\bvolume{52}(\bissue{15}),
\bfpage{2025}--\blpage{115926}
(\byear{2025})
\end{barticle}
\endbibitem

\bibitem[\protect\citeauthoryear{Thompson et~al.}{2004}]{thompson2004explicit}
\begin{barticle}
\bauthor{\bsnm{Thompson}, \binits{G.}},
\bauthor{\bsnm{Rasmussen}, \binits{R.M.}},
\bauthor{\bsnm{Manning}, \binits{K.}}:
\batitle{Explicit forecasts of winter precipitation using an improved bulk microphysics scheme. part i: Description and sensitivity analysis}.
\bjtitle{Monthly Weather Review}
\bvolume{132}(\bissue{2}),
\bfpage{519}--\blpage{542}
(\byear{2004})
\end{barticle}
\endbibitem

\bibitem[\protect\citeauthoryear{Zheng et~al.}{2016}]{zheng2016improving}
\begin{barticle}
\bauthor{\bsnm{Zheng}, \binits{Y.}},
\bauthor{\bsnm{Alapaty}, \binits{K.}},
\bauthor{\bsnm{Herwehe}, \binits{J.A.}},
\bauthor{\bsnm{Del~Genio}, \binits{A.D.}},
\bauthor{\bsnm{Niyogi}, \binits{D.}}:
\batitle{Improving high-resolution weather forecasts using the weather research and forecasting (wrf) model with an updated kain--fritsch scheme}.
\bjtitle{Monthly Weather Review}
\bvolume{144}(\bissue{3}),
\bfpage{833}--\blpage{860}
(\byear{2016})
\end{barticle}
\endbibitem

\bibitem[\protect\citeauthoryear{Iacono et~al.}{2008}]{iacono2008radiative}
\begin{botherref}
\oauthor{\bsnm{Iacono}, \binits{M.J.}},
\oauthor{\bsnm{Delamere}, \binits{J.S.}},
\oauthor{\bsnm{Mlawer}, \binits{E.J.}},
\oauthor{\bsnm{Shephard}, \binits{M.W.}},
\oauthor{\bsnm{Clough}, \binits{S.A.}},
\oauthor{\bsnm{Collins}, \binits{W.D.}}:
Radiative forcing by long-lived greenhouse gases: Calculations with the aer radiative transfer models.
Journal of Geophysical Research: Atmospheres
\textbf{113}(D13)
(2008)
\end{botherref}
\endbibitem

\bibitem[\protect\citeauthoryear{Ek et~al.}{2003}]{ek2003implementation}
\begin{botherref}
\oauthor{\bsnm{Ek}, \binits{M.}},
\oauthor{\bsnm{Mitchell}, \binits{K.}},
\oauthor{\bsnm{Lin}, \binits{Y.}},
\oauthor{\bsnm{Rogers}, \binits{E.}},
\oauthor{\bsnm{Grunmann}, \binits{P.}},
\oauthor{\bsnm{Koren}, \binits{V.}},
\oauthor{\bsnm{Gayno}, \binits{G.}},
\oauthor{\bsnm{Tarpley}, \binits{J.}}:
Implementation of noah land surface model advances in the national centers for environmental prediction operational mesoscale eta model.
Journal of Geophysical Research: Atmospheres
\textbf{108}(D22)
(2003)
\end{botherref}
\endbibitem

\bibitem[\protect\citeauthoryear{Hu et~al.}{2013}]{hu2013evaluation}
\begin{barticle}
\bauthor{\bsnm{Hu}, \binits{X.-M.}},
\bauthor{\bsnm{Klein}, \binits{P.M.}},
\bauthor{\bsnm{Xue}, \binits{M.}}:
\batitle{Evaluation of the updated ysu planetary boundary layer scheme within wrf for wind resource and air quality assessments}.
\bjtitle{Journal of Geophysical Research: Atmospheres}
\bvolume{118}(\bissue{18}),
\bfpage{10}--\blpage{490}
(\byear{2013})
\end{barticle}
\endbibitem

\bibitem[\protect\citeauthoryear{der Grijn}{2002}]{der2002tropical}
\begin{barticle}
\bauthor{\bsnm{Grijn}, \binits{V.}}:
\batitle{Tropical cyclone forecasting at ecmwf: New products and validation}.
\bjtitle{ECMWF Tech. Memo.}
\bvolume{386},
\bfpage{1}
(\byear{2002})
\end{barticle}
\endbibitem

\bibitem[\protect\citeauthoryear{Wilks}{2011}]{Wilks2011}
\begin{bbook}
\bauthor{\bsnm{Wilks}, \binits{D.S.}}:
\bbtitle{Statistical Methods in the Atmospheric Sciences},
\bedition{3rd} edn.
\bpublisher{Academic Press},
\blocation{Oxford}
(\byear{2011})
\end{bbook}
\endbibitem

\bibitem[\protect\citeauthoryear{Tustison et~al.}{2001}]{Tustison2001}
\begin{barticle}
\bauthor{\bsnm{Tustison}, \binits{B.}},
\bauthor{\bsnm{Harris}, \binits{D.}},
\bauthor{\bsnm{Foufoula-Georgiou}, \binits{E.}}:
\batitle{Scale issues in verification of precipitation forecasts}.
\bjtitle{Journal of Geophysical Research: Atmospheres}
\bvolume{106}(\bissue{D11}),
\bfpage{11775}--\blpage{11784}
(\byear{2001})
\doiurl{10.1029/2001JD900066}
\end{barticle}
\endbibitem

\bibitem[\protect\citeauthoryear{Accadia et~al.}{2003}]{accadia2003sensitivity}
\begin{barticle}
\bauthor{\bsnm{Accadia}, \binits{C.}},
\bauthor{\bsnm{Mariani}, \binits{S.}},
\bauthor{\bsnm{Casaioli}, \binits{M.}},
\bauthor{\bsnm{Lavagnini}, \binits{A.}},
\bauthor{\bsnm{Speranza}, \binits{A.}}:
\batitle{Sensitivity of precipitation forecast skill scores to bilinear interpolation and a simple nearest-neighbor average method on high-resolution verification grids}.
\bjtitle{Weather and forecasting}
\bvolume{18}(\bissue{5}),
\bfpage{918}--\blpage{932}
(\byear{2003})
\end{barticle}
\endbibitem

\bibitem[\protect\citeauthoryear{Accadia et~al.}{2005}]{accadia2005verification}
\begin{barticle}
\bauthor{\bsnm{Accadia}, \binits{C.}},
\bauthor{\bsnm{Mariani}, \binits{S.}},
\bauthor{\bsnm{Casaioli}, \binits{M.}},
\bauthor{\bsnm{Lavagnini}, \binits{A.}},
\bauthor{\bsnm{Speranza}, \binits{A.}}:
\batitle{Verification of precipitation forecasts from two limited-area models over italy and comparison with ecmwf forecasts using a resampling technique}.
\bjtitle{Weather and Forecasting}
\bvolume{20}(\bissue{3}),
\bfpage{276}--\blpage{300}
(\byear{2005})
\end{barticle}
\endbibitem

\bibitem[\protect\citeauthoryear{Lu et~al.}{2022}]{lu2022dpmsolverpp}
\begin{botherref}
\oauthor{\bsnm{Lu}, \binits{C.}},
\oauthor{\bsnm{Zhou}, \binits{Y.}},
\oauthor{\bsnm{Bao}, \binits{F.}},
\oauthor{\bsnm{Chen}, \binits{J.}},
\oauthor{\bsnm{Li}, \binits{C.}},
\oauthor{\bsnm{Zhu}, \binits{J.}}:
Dpm-solver++: Fast high-order solver for diffusion probabilistic models.
arXiv preprint arXiv:2211.01095
(2022)
\end{botherref}
\endbibitem

\bibitem[\protect\citeauthoryear{Guo}{2026}]{guo_2026_19382719}
\begin{botherref}
\oauthor{\bsnm{Guo}, \binits{S.}}:
FuXi-TC.
Zenodo
(2026).
\doiurl{10.5281/zenodo.19382719} .
\url{https://doi.org/10.5281/zenodo.19382719}
\end{botherref}
\endbibitem

\end{thebibliography}

\end{document}


\title{Supplementary Information: FuXi-TC}


\author[2]{\fnm{Shan} \sur{Guo}}\email{guoshan@sais.org.cn}
\equalcont{These authors contributed equally to this work.}

\author[1,2]{\fnm{Lei} \sur{Chen}}\email{cltpys@163.com}
\equalcont{These authors contributed equally to this work.}

\author[1,2]{\fnm{Yangyang} \sur{Zhao}}\email{zhaoyangyang@sais.org.cn}

\author[1]{\fnm{Yuetan} \sur{Lin}}\email{linyuetan@sais.org.cn}

\author[3,4]{\fnm{Zeyi} \sur{Niu}}\email{niuzy@typhoon.org.cn}

\author[3]{\fnm{Xinyan} \sur{Zhang}}\email{zhangxy@typhoon.org.cn}

\author[3,4]{\fnm{Ziyao} \sur{Sun}}\email{sunzy@typhoon.org.cn}

\author*[1,2]{\fnm{Xiaohui} \sur{Zhong}}\email{x7zhong@gmail.com}

\author*[1,2]{\fnm{Hao} \sur{Li}}\email{lihao$\_$lh@fudan.edu.cn}

\affil[1]{\orgdiv{Artificial Intelligence Innovation and Incubation Institute}, \orgname{Fudan University}, \orgaddress{\city{Shanghai}, \postcode{200433}, \country{China}}}

\affil[2]{\orgname{Shanghai Academy of Artificial Intelligence for Science}, \orgaddress{\city{Shanghai}, \postcode{200232}, \country{China}}}

\affil[3]{\orgdiv{Department of Atmospheric and Oceanic Sciences, Institute of Atmospheric Sciences}, \orgname{Fudan University}, \orgaddress{\city{Shanghai}, \postcode{200433}, \country{China}}}

\affil[4]{\orgdiv{Shanghai Typhoon Institute}, \orgname{China Meteorological Administration}, \orgaddress{\city{Shanghai}, \postcode{200030}, \country{China}}}


\maketitle





\section{Tropical cyclone cases}

Table~\ref{tab:tc_summary} lists the tropical cyclone (TC) cases evaluated in both the Western North Pacific (WNP) and North Atlantic (NA) basins. 
For each storm, the storm identifier (SID) from International Best Track Archive for Climate Stewardship (IBTrACS) and the corresponding first and last forecast initialization times (UTC) are provided. All cases occurred in 2024.
The inclusion of 17 WNP and 15 NA storms ensures coverage of diverse dynamical regimes and basin-dependent characteristics. 
By evaluating multiple initialization times for each TC, we systematically assess forecast performance across different lead times and storm development stages, thereby enabling a comprehensive validation of the proposed framework.

\begin{table}[htbp]
    \centering
    \caption{List of TC names, TC SID and their first and last initialization times (in UTC) in WNP and NA Basins evaluated in this study. All dates refer to the year 2024.}
    \label{tab:tc_summary}
    \begin{tabular}{cllccc}
        \toprule
        \textbf{Basin} & \textbf{TC Name} & \textbf{TC SID} & \textbf{First Init Time} & \textbf{Last Init Time} & \textbf{Init Times} \\
        \midrule
        \multirow{17}{*}{WNP} 
        & GAEMI    & 2024201N12133 & 20 Jul, 1200 & 25 Jul, 1200 & 11 \\
        & PRAPIROON& 2024201N14118 & 21 Jul, 1200 & 22 Jul, 1200 & 3  \\
        & MARIA    & 2024218N24141 & 08 Aug, 0000 & 12 Aug, 0000 & 9  \\
        & SON-TINH & 2024224N27154 & 12 Aug, 0000 & 12 Aug, 0000 & 1  \\
        & AMPIL    & 2024225N22135 & 13 Aug, 0000 & 18 Aug, 0000 & 11 \\
        & WUKONG   & 2024225N24147 & 13 Aug, 1200 & 13 Aug, 1200 & 1  \\
        & JONGDARI & 2024231N24126 & 19 Aug, 0000 & 19 Aug, 1200 & 2  \\
        & SHANSHAN & 2024233N15148 & 22 Aug, 0000 & 29 Aug, 0000 & 15 \\
        & YAGI     & 2024245N12129 & 01 Sep, 1200 & 07 Sep, 1200 & 13 \\
        & LEEPI    & 2024246N22147 & 05 Sep, 1200 & 05 Sep, 1200 & 1  \\
        & BEBINCA  & 2024254N10148 & 10 Sep, 1200 & 16 Sep, 1200 & 13 \\
        & PULASAN  & 2024259N12145 & 15 Sep, 1200 & 20 Sep, 1200 & 11 \\
        & JEBI     & 2024270N15149 & 27 Sep, 1200 & 01 Oct, 1200 & 8  \\
        & KRATHON  & 2024271N22127 & 28 Sep, 0000 & 02 Oct, 1200 & 10 \\
        & BARIJAT  & 2024279N12148 & 09 Oct, 1200 & 10 Oct, 1200 & 3  \\
        & TRAMI    & 2024293N13141 & 22 Oct, 0000 & 27 Oct, 0000 & 11 \\
        & KONG-REY & 2024298N13150 & 25 Oct, 0000 & 31 Oct, 1200 & 13 \\
        \midrule
        \multirow{15}{*}{NA} 
        & BERYL    & 2024181N09320 & 29 Jun, 0000 & 08 Jul, 1200 & 20 \\
        & DEBBY    & 2024216N20284 & 04 Aug, 0000 & 09 Aug, 1200 & 11 \\
        & ERNESTO  & 2024225N14313 & 13 Aug, 0000 & 20 Aug, 0000 & 15 \\
        & FRANCINE & 2024253N21266 & 10 Sep, 0000 & 12 Sep, 1200 & 6  \\
        & GORDON   & 2024256N16332 & 13 Sep, 1200 & 17 Sep, 0000 & 8  \\
        & HELENE   & 2024268N17278 & 25 Sep, 0000 & 28 Sep, 0000 & 7  \\
        & ISAAC    & 2024269N39302 & 26 Sep, 1200 & 30 Sep, 1200 & 9  \\
        & JOYCE    & 2024271N17319 & 27 Sep, 1200 & 29 Sep, 1200 & 5  \\
        & KIRK     & 2024274N14328 & 01 Oct, 0000 & 08 Oct, 0000 & 15 \\
        & LESLIE   & 2024275N10336 & 03 Oct, 0000 & 12 Oct, 0000 & 19 \\
        & MILTON   & 2024279N21265 & 06 Oct, 0000 & 11 Oct, 0000 & 11 \\
        & NADINE   & 2024293N17275 & 19 Oct, 1200 & 20 Oct, 0000 & 2  \\
        & OSCAR    & 2024293N21294 & 20 Oct, 0000 & 22 Oct, 0000 & 5  \\
        & PATTY    & 2024306N39319 & 02 Nov, 1200 & 04 Nov, 0000 & 3  \\
        & RAFAEL   & 2024309N13283 & 05 Nov, 0000 & 10 Nov, 0000 & 11 \\
        \bottomrule
    \end{tabular}
\end{table}

\section{Comparison between WRF-0.1\texorpdfstring{$^\circ$}{°} and WRF-0.25\texorpdfstring{$^\circ$}{°}}
\label{compare_wrf0.1_wrf0.25}
\begin{figure}[t]
    \centering
    \includegraphics[width=\linewidth]{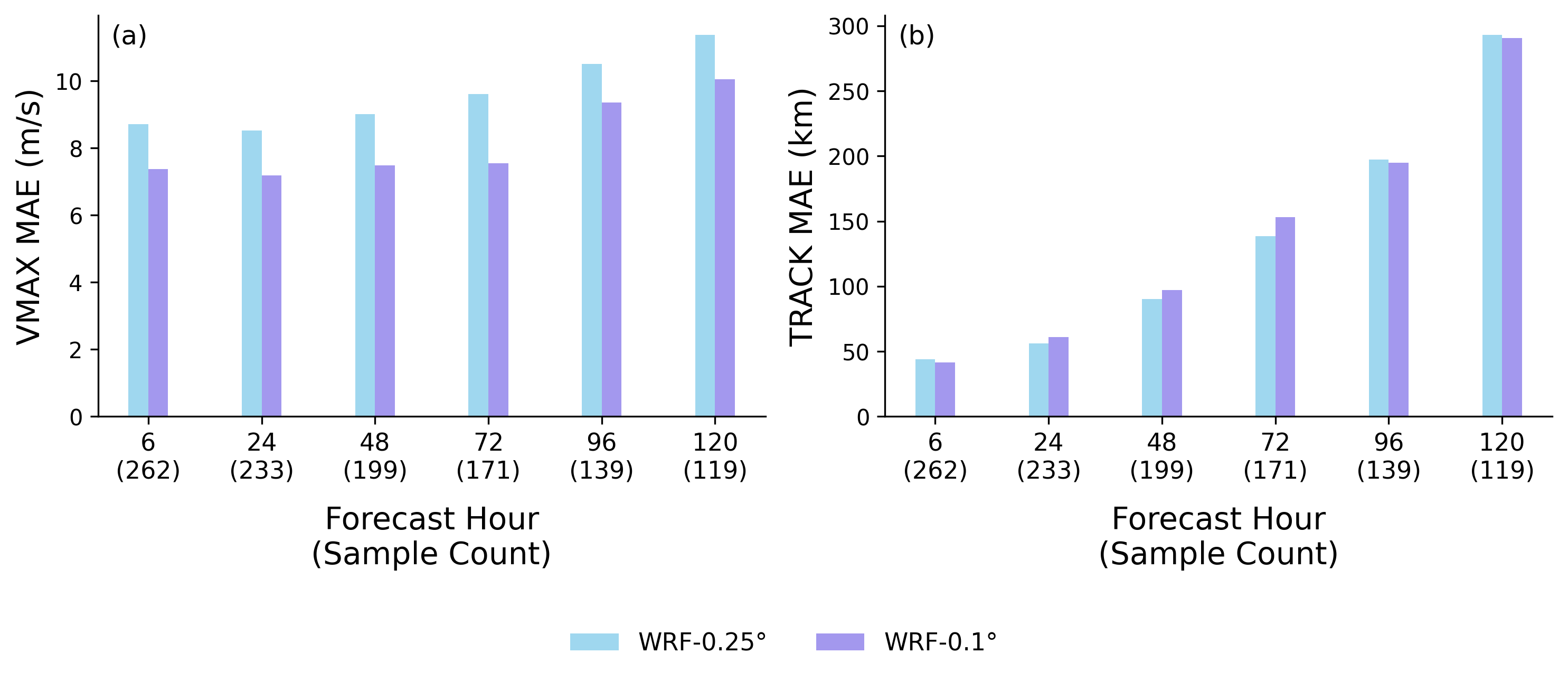}
    \caption{Comparison of TC intensity and track forecasts between WRF simulations at different resolutions with IBTrACS dataset as ground truth. Mean absolute error (MAE) of maximum sustained wind and tracks with forecast lead time for WRF-\(0.25^\circ\) (blue columns), WRF-\(0.1^\circ\) (purple columns).}
    \label{wrf_compare}            
\end{figure}

We evaluated the model sensitivity to horizontal resolution by comparing simulations at \(0.25^\circ\) and \(0.1^\circ\) using identical physical configurations from April to October in 2023. Supplementary Figure \ref{wrf_compare} illustrates while track forecast accuracy remains similar between the two resolutions, the WRF-\(0.1^\circ\) configuration demonstrates significantly superior performance in predicting TC intensity compared to WRF-\(0.25^\circ\).

\FloatBarrier

\section{Spatial distribution of observed precipitation event counts}

\begin{figure}[t]
    \centering
    \includegraphics[width=\linewidth]{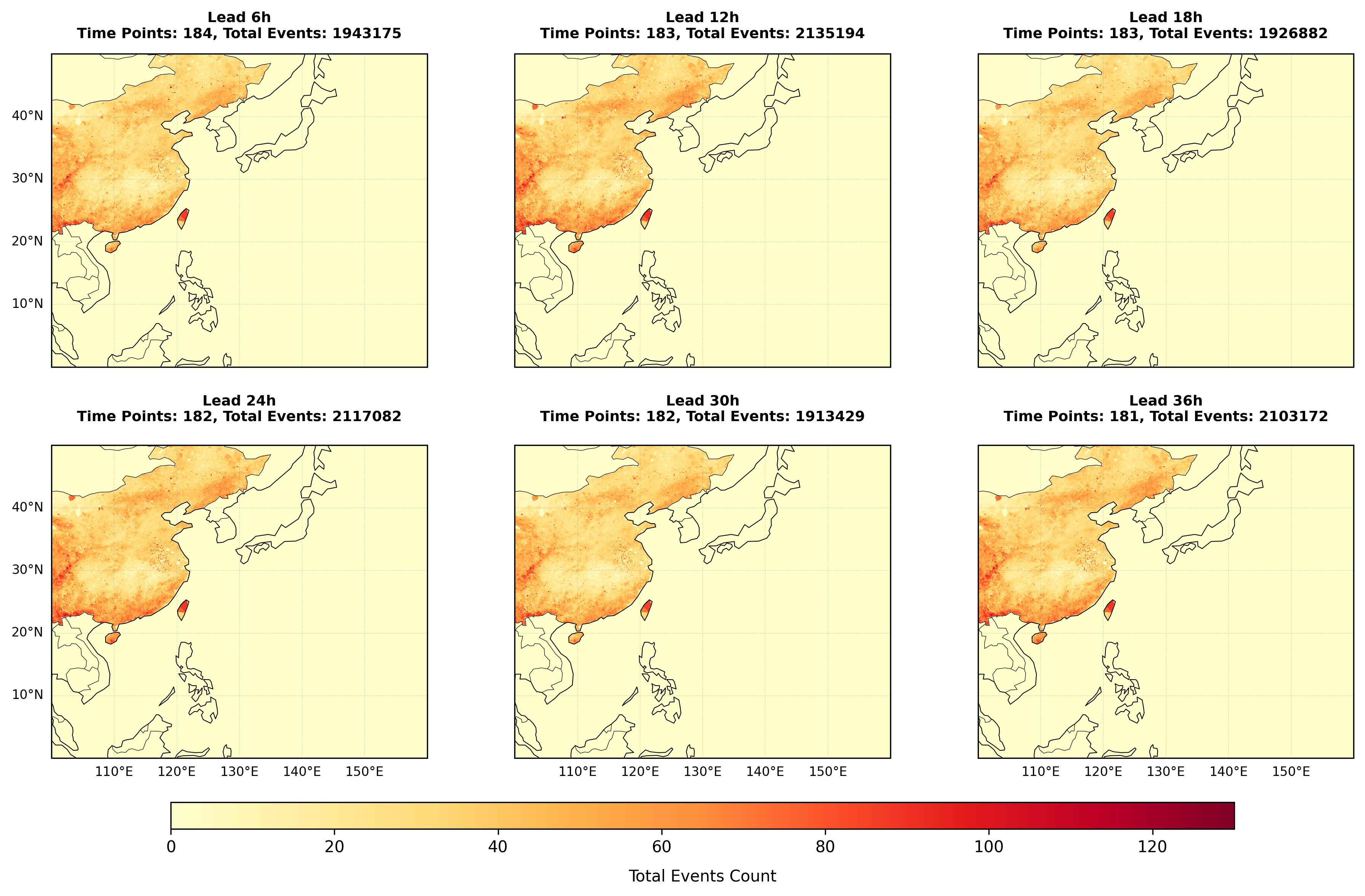}
    \caption{Spatial distribution of observed precipitation event counts with a threshold of 0.1 mm by forecast lead time during July–September 2024.}
    \label{total_events}            
\end{figure}

Figure \ref{total_events} illustrates the spatial distribution of Automatic Weather Station (AWS)-based precipitation event statistics for rainfall exceeding 0.1 mm from July to September 2024 and their temporal variations across different forecast lead times. 
Given that the experiments were initialized at 00 and 12 UTC, this figure reveals that the volume of precipitation data received from AWS at 00 and 12 UTC is significantly larger than that available at 06 and 18 UTC.
  
\FloatBarrier

\section{Prediction of 2024 Typhoon Bebinca}
\label{bebinca_cases}
\begin{figure}[t]
    \centering
    \includegraphics[width=\linewidth]{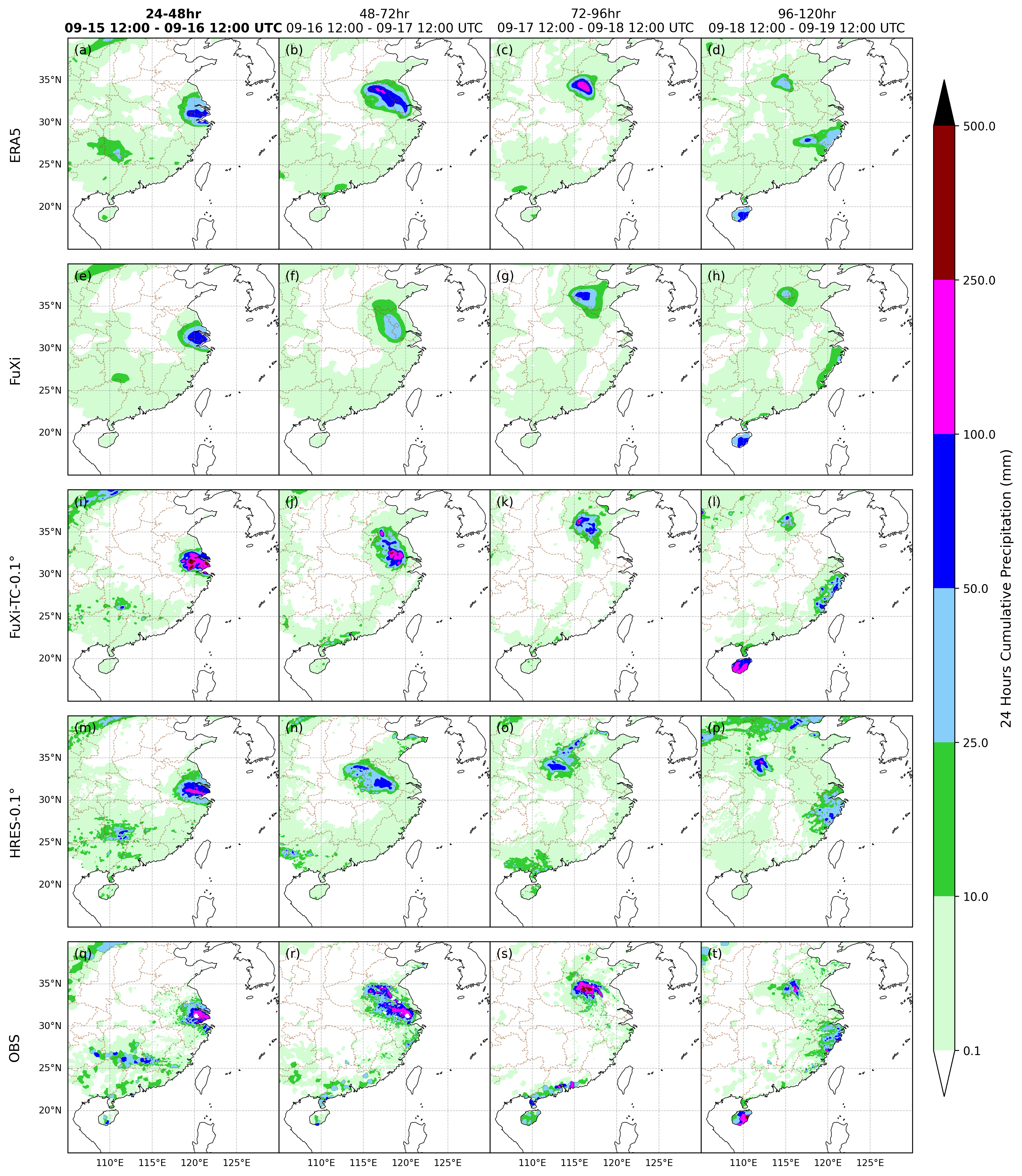}
    \caption{Comparison of total precipitation (TP) for Typhoon Bebinca and Tropical Storm Pulasan forecasts. Comparison of snapshot examples of 24-hour accumulated TP among ERA5 (first row), FuXi (second row), FuXi-TC-\(0.1^\circ\) (third row), HRES-\(0.1^\circ\) (fourth row) and the AWS observations (fifth row) mainland China for 24-48 (first column), 48-72 (second column), 72-96 (third column) and 96-120 (fourth column) hours forecasts with the initial forecasting time at 12:00 UTC 14 September 2024.}
    \label{bebinca_precip}            
\end{figure}

Supplementary Figure \ref{bebinca_precip} shows the 24-hour accumulated precipitation observation from AWS and forecasts initialized at 12 UTC on 14 September 2024. The comparison covers forecast lead times of 24-48, 48-72, 72-96 and 96-120 hours. These precipitation events encompass the rainfall over land induced by Typhoon Bebinca and Tropical Storm Pulasan. Notably, Bebinca was the strongest typhoon to make landfall in Shanghai since 1949, followed by Tropical Storm Pulasan, which made a subsequent landfall in Shanghai just four days later \cite{wu2025extreme}. According to observations, Bebinca triggered heavy to torrential rainfall extending from the eastern coast to Central China, while Pulasan produced heavy rainfall with localized torrential downpours along the southeastern coast. Regarding models' performance, the FuXi model generally underestimates precipitation intensity. HRES-\(0.1^\circ\) exhibits spatial displacement during the 48–72 h and 72–96 h intervals. However, FuXi-TC-\(0.1^\circ\) inherits the precise spatial positioning of FuXi while demonstrating significant improvements in capturing precipitation intensity.

\FloatBarrier

\begin{figure}[t]
    \centering
    \includegraphics[width=\linewidth]{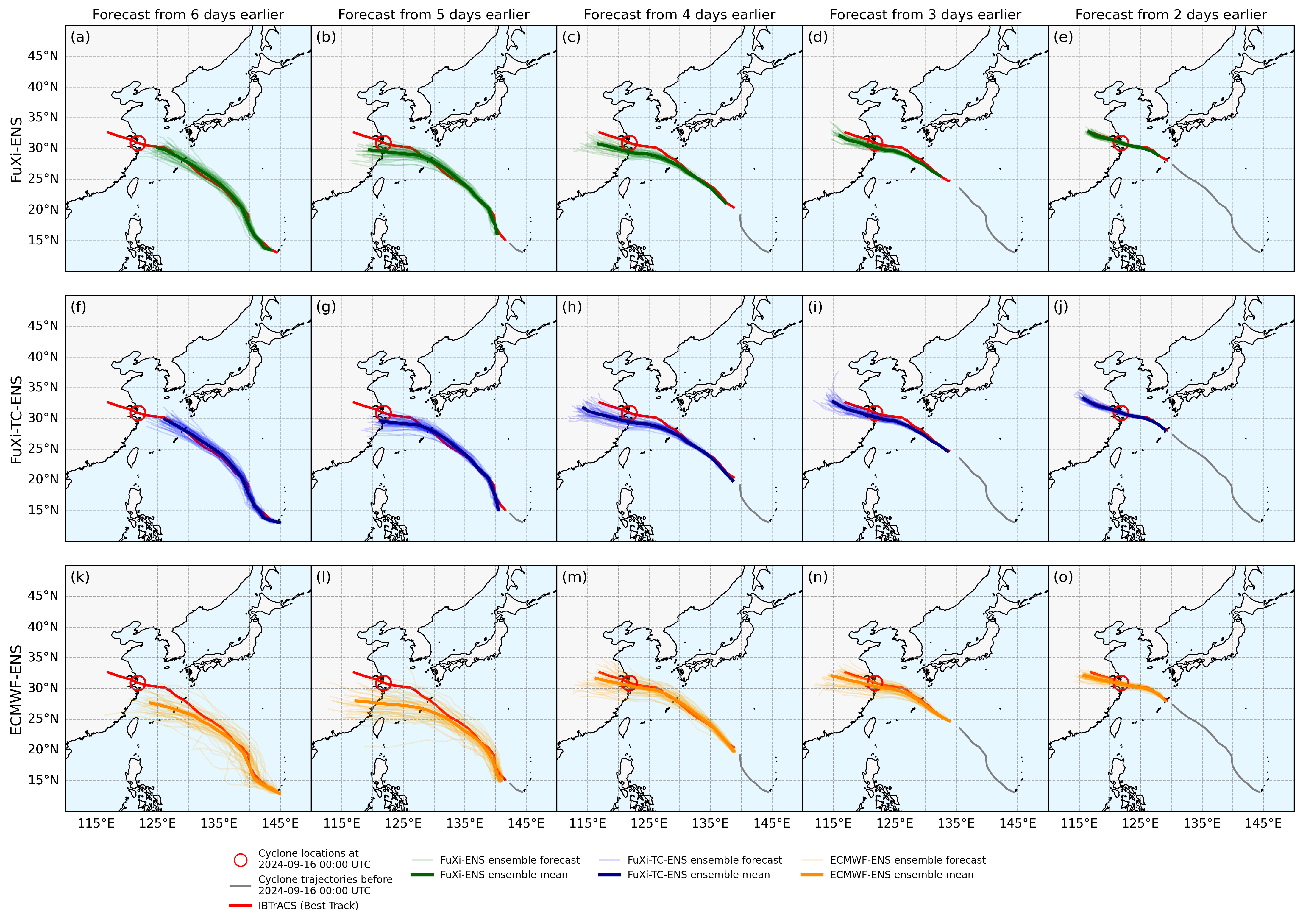}
    \caption{Tracking results for Typhoon Bebinca from ERA5 (red line) and ensemble forecasts generated by FuXi-ENS (first row, green line), FuXi-TC-ENS (second row, blue line) and ECMWF-ENS (third row, orange line). The columns show forecasts initialized at 12:00 UTC on 10, 11, 12, 13, and 14 September 2024 (from left to right), corresponding to 6, 5, 4, 3, and 2 days before landfall, respectively. The gray line represents the observed track prior to each forecast initialization time, and the red circle indicates the landfall location.}
    \label{track_ens}   
\end{figure}

Supplementary Figure \ref{track_ens} presents a comparison of forecast tracks for Typhoon Bebinca among FuXi-ENS, FuXi-TC-ENS, and ECMWF-ENS across different initialization times. Notably, the differences between the ensemble mean tracks of FuXi-TC-ENS and FuXi-ENS are negligible. This is attributed to the fact that the post-processing framework primarily adjusts typhoon intensity and exerts minimal influence on the track. Regarding TC track forecasts, ECMWF-ENS exhibits slightly inferior performance in this case study compared to both FuXi-ENS and FuXi-TC-ENS, particularly for forecasts initialized 5 to 6 days in advance.

\FloatBarrier

\section{Overall precipitation evaluation in WNP}
\label{WNP_precipiation_evaluation}

\begin{figure}[t]
    \centering
    \includegraphics[width=\linewidth]{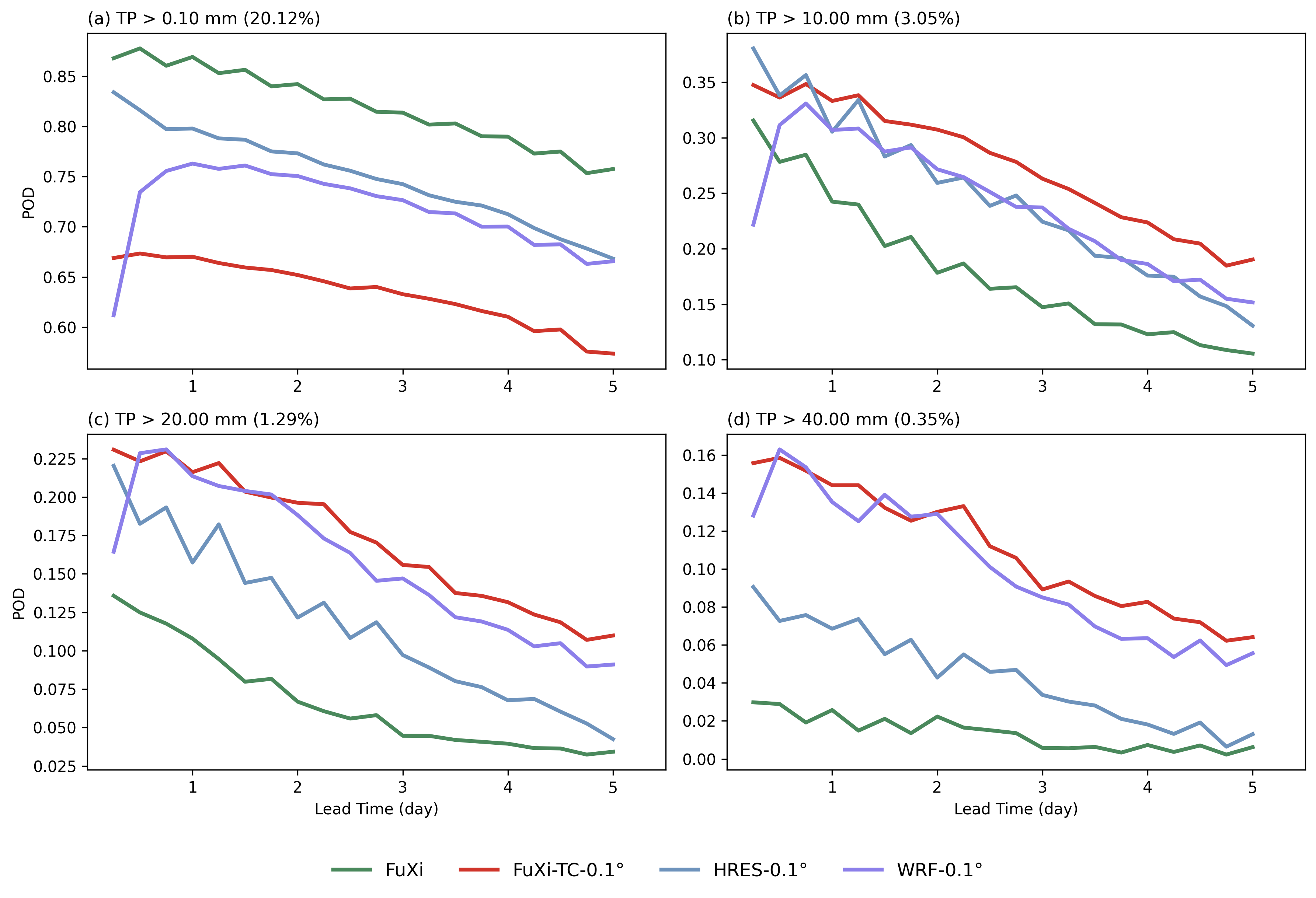}
    \caption{Comparison of POD for TP. Comparison for the FuXi(green lines), FuXi-TC-\(0.1^\circ\)(red lines), HRES-\(0.1^\circ\)(blue lines) and WRF-\(0.1^\circ\)(purple lines) in predicting 6-hour precipitation for 0.1, 10, 20, 40 mm thresholds in (a-d) respectively. The ratios of extreme cases relative to the entire test set are indicated in parentheses. All the forecast data are evaluated against the Automatic Weather Station observations in period 1 July to 30 September 2024.}
    \label{pod}   
\end{figure}

\begin{figure}[t]
    \centering
    \includegraphics[width=\linewidth]{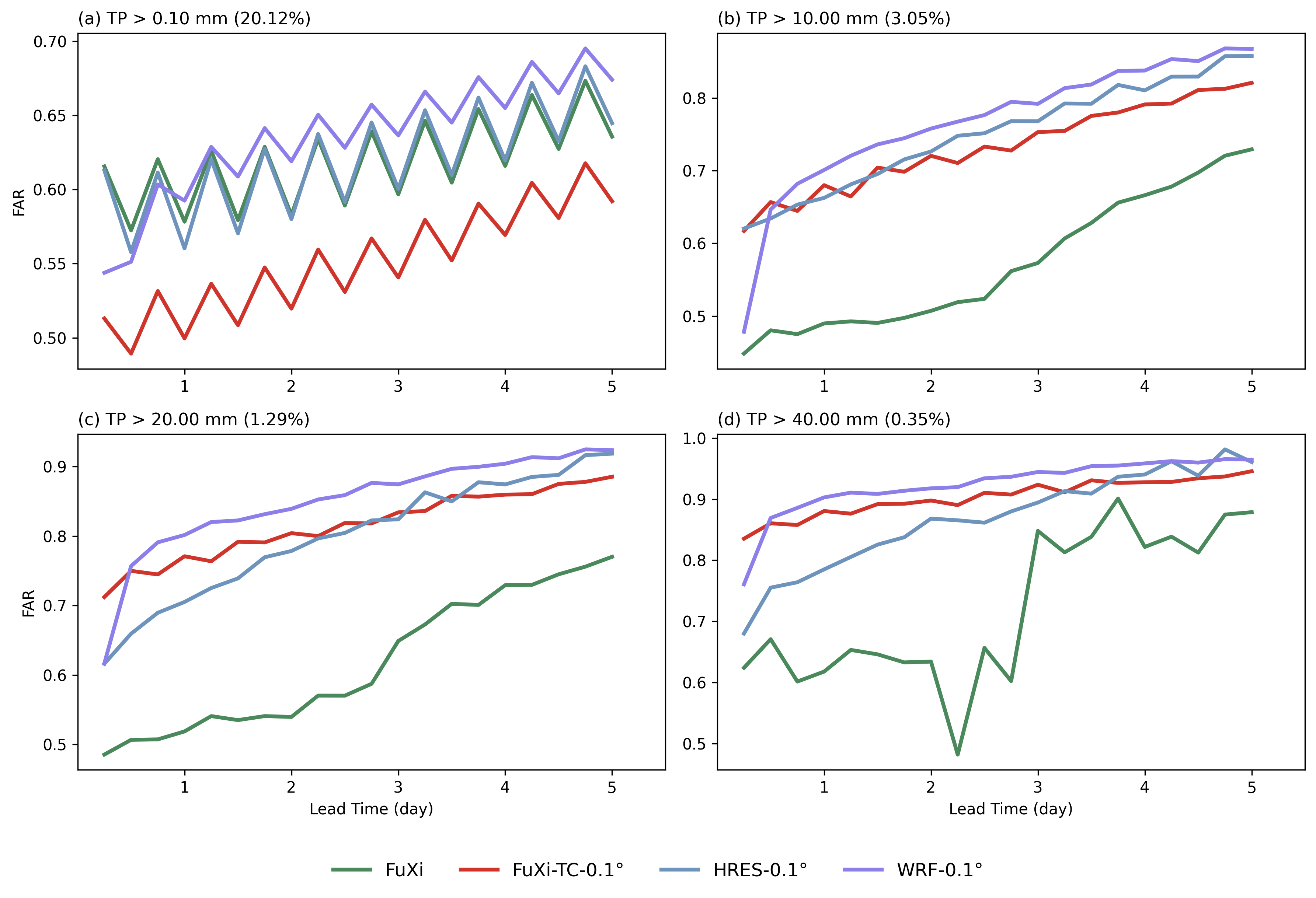}
    \caption{Comparison of FAR for TP. Comparison for the FuXi (green lines), FuXi-TC-\(0.1^\circ\) (red lines), HRES-\(0.1^\circ\) (blue lines) and WRF-\(0.1^\circ\) (purple lines) in predicting 6-hour precipitation for 0.1, 10, 20, 40 mm thresholds in (a-d) respectively. The ratios of extreme cases relative to the entire test set are indicated in parentheses. All the forecast data are evaluated against the Automatic Weather Station observations in period 1 July to 30 September 2024.}
    \label{far}   
\end{figure}

Supplementary Figure \ref{pod} and \ref{far} present the time series of Probability of Detection (POD) and False Alarm Ratio (FAR) for 6-hour accumulated TP across varying thresholds from 0.1 mm to 40 mm. POD reflects the model's sensitivity to observed precipitation events, while FAR indicates the propensity for false alarms. In terms of model performance, FuXi performs well at the lowest threshold. However, at higher thresholds, both its POD and FAR decrease significantly, a trend attributed to the model's general tendency to underestimate precipitation intensity. Notably, in the evaluation for TP at 40 mm threshold, FuXi's FAR exhibits large fluctuations, likely due to the insufficient sample size of extreme precipitation events predicted by FuXi. Comparatively, HRES-\(0.1^\circ\) demonstrates a relatively stable trade-off between POD and FAR across thresholds. Both WRF-\(0.1^\circ\) and FuXi-TC-\(0.1^\circ\) show superiority at higher precipitation thresholds. WRF-\(0.1^\circ\) exhibits a higher FAR at thresholds above 0.1 mm, possibly due to a systematic wet bias in the numerical model. During the initial forecast period, both POD and FAR are relatively low for WRF-\(0.1^\circ\) , reflecting the spin-up process during which precipitation is not fully developed.

\FloatBarrier

\FloatBarrier

\section{Energy spectra}
\label{energy_spectra}

\begin{figure}[t]
    \centering
    \includegraphics[width=\linewidth]{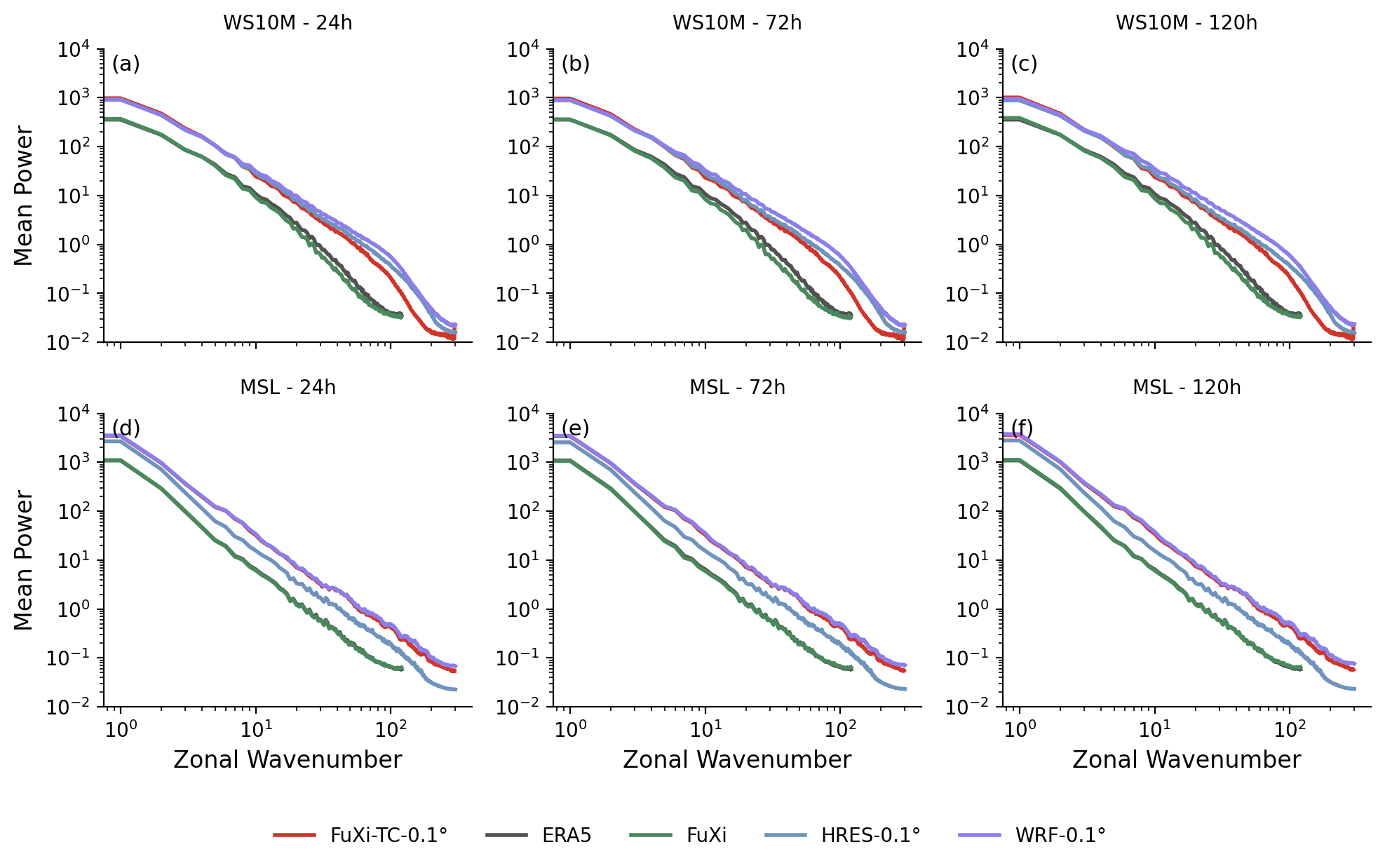}
    \caption{Energy spectra of FuXi-TC-\(0.1^\circ\)(red lines), ERA5(black lines), FuXi(green lines), HRES-\(0.1^\circ\)(blue lines) and WRF-\(0.1^\circ\)(purple lines). The figure includes WS10M (first row) and MSL (second row). The spectra are shown at forecast lead times of 24 hours (first column), 72 hours (second column) and 120 hours (third column), using the dataset from July to October 2024.}
    \label{spectra}            
\end{figure}

We employed kinetic energy spectra to evaluate the effective resolution and smoothness of the model outputs. By decomposing the kinetic energy into spectral components, we assessed the model's ability to preserve variance across different scales. Generally, spectral energy exhibits a power-law decay as wavelengths decrease, reflecting the energy cascade from organized large-scale circulation to smaller-scale turbulent processes. Supplementary Figure \ref{spectra} presents energy spectra within the WNP domain for 10-meter wind speed (WS10M) and  mean sea level pressure (MSL) across 24, 72, 120 hours forecast lead times. The \(0.1^\circ\)-resolution models exhibit distinct spectral characteristics compared to their \(0.25^\circ\) counterparts. FuXi achieves remarkable alignment with European Centre for Medium-Range Weather Forecasts Reanalysis v5 (ERA5) across all scales and surface variables. Notably, FuXi-TC-\(0.1^\circ\) significantly outperforms the FuXi. At smaller wavelengths, FuXi-TC-\(0.1^\circ\) initially aligns with WRF-\(0.1^\circ\) but shows slightly reduced energy with increasing wavenumber, suggesting smoother forecasts and a loss of fine-scale details for WS10M. However, for MSL,  FuXi-TC-\(0.1^\circ\) indicates superior consistency with WRF-\(0.1^\circ\), suggesting that the model is more capable of capturing small-scale features and details for this variable.

\FloatBarrier

\section{Overall MSL evaluation in WNP and NA}
\label{WNP_NA_MSL_evaluation}
\begin{figure}[t]
    \centering
    \includegraphics[width=\linewidth]{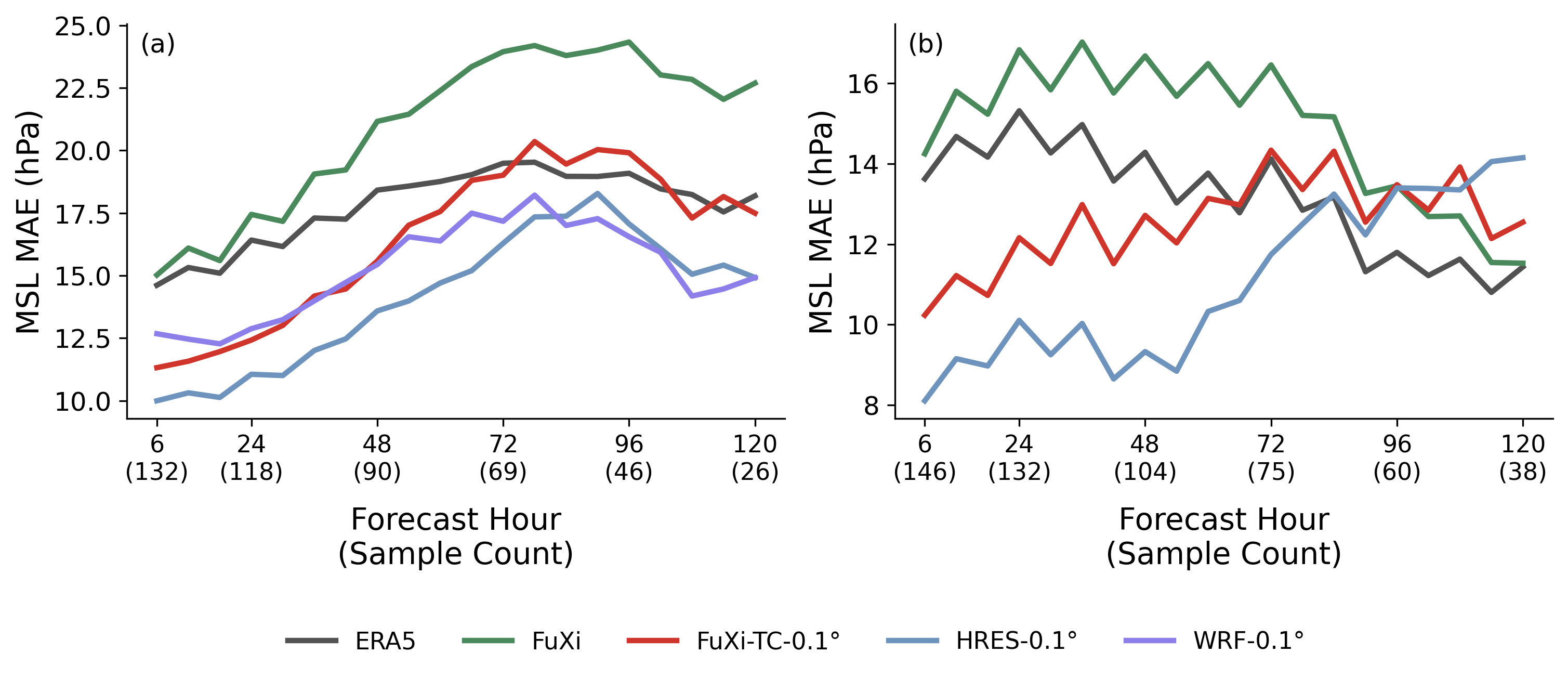}
    \caption{Comparison of the average MAE for MSL in WNP (a) and NA (b) with forecast lead time for ERA5 (black lines), FuXi (green lines), FuXi-TC-\(0.1^\circ\) (red lines), HRES-\(0.1^\circ\) (blue lines) and WRF-\(0.1^\circ\) (purple lines). 
    All the forecast data are evaluated against the IBTrACS dataset.
    The evaluation covers forecasts for 17 TCs in WNP from July to October 2024, and 15 TCs in NA from June to December.}
    \label{msl_mae}   
\end{figure}

Supplementary Figure \ref{msl_mae} presents the overall MAE of the MSL for TCs in both the WNP and NA basins. In the WNP, FuXi-TC-\(0.1^\circ\) demonstrates a significant improvement in MSL accuracy over the original FuXi model across all lead times. For lead times within the first 48 hours, its performance is comparable to that of WRF-\(0.1^\circ\). However, since WRF-\(0.1^\circ\) exhibits a higher MAE than HRES-\(0.1^\circ\), HRES-\(0.1^\circ\) remains the superior performer overall.  The improvement in MSL accuracy relative to FuXi remains robust in different regions, which demonstrates that the FuXi-TC framework can be directly applied to the NA basin without retraining using WRF-\(0.1^\circ\) data.

\FloatBarrier

\section{Wind speed prediction comparison of tropical cyclones cases in NA}
\label{ws_predicition_in_NA}

\begin{figure}[t]
    \centering
    \includegraphics[width=\linewidth]{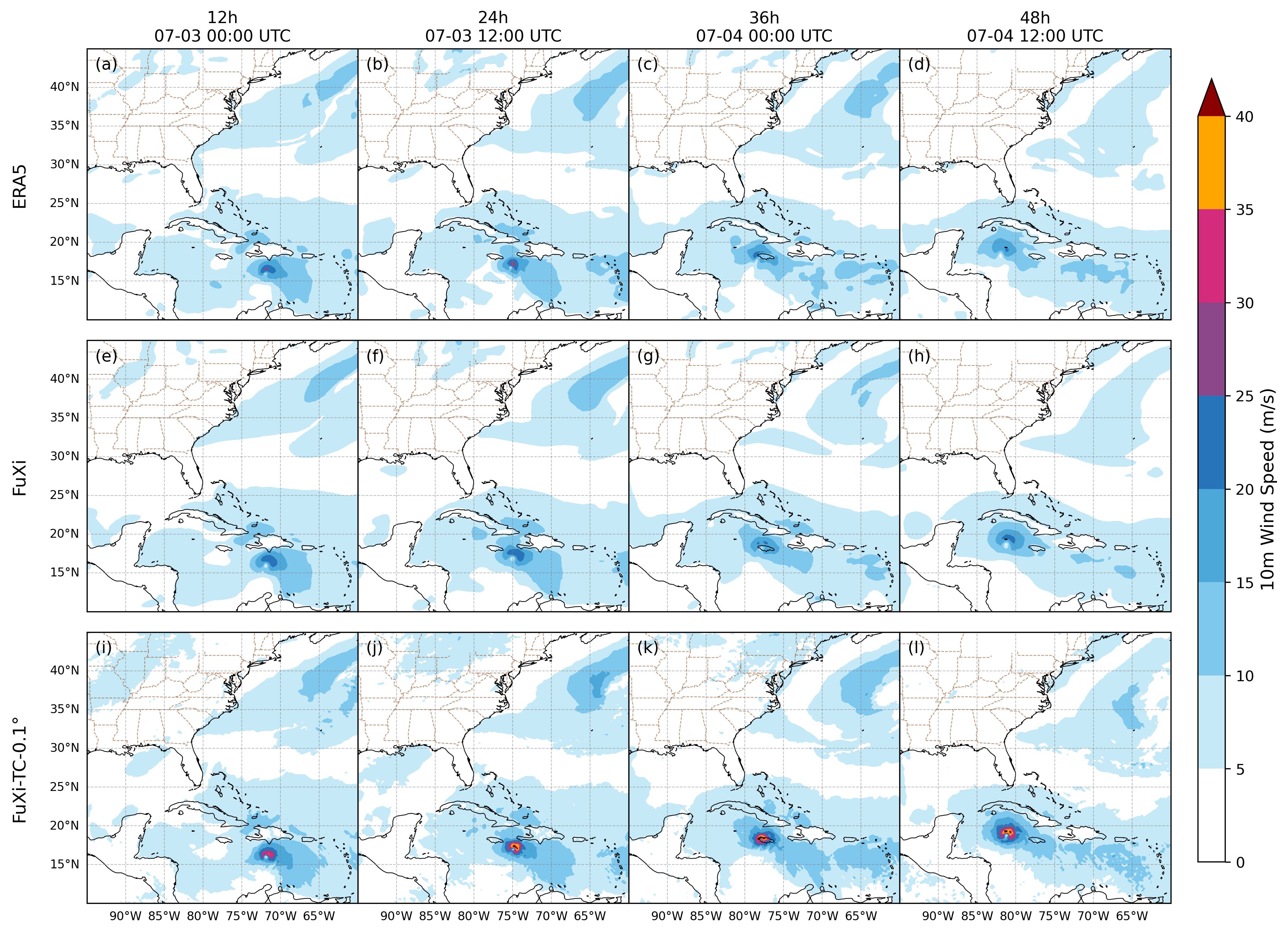}
    \caption{Comparison of WS10M for Hurricane Beryl forecast. Comparison of snapshot examples of WS10M among ERA5 (first row), FuXi (second row) and FuXi-TC-\(0.1^\circ\) (third row) for 12 (first column), 24 (second column), 36 (third column) and 48 (fourth row) hours forecasts with the initial forecasting time at 12:00 UTC, 2 July, 2024.}
    \label{atlantic_beryl}
\end{figure}

\begin{figure}[t]
    \centering
    \includegraphics[width=\linewidth]{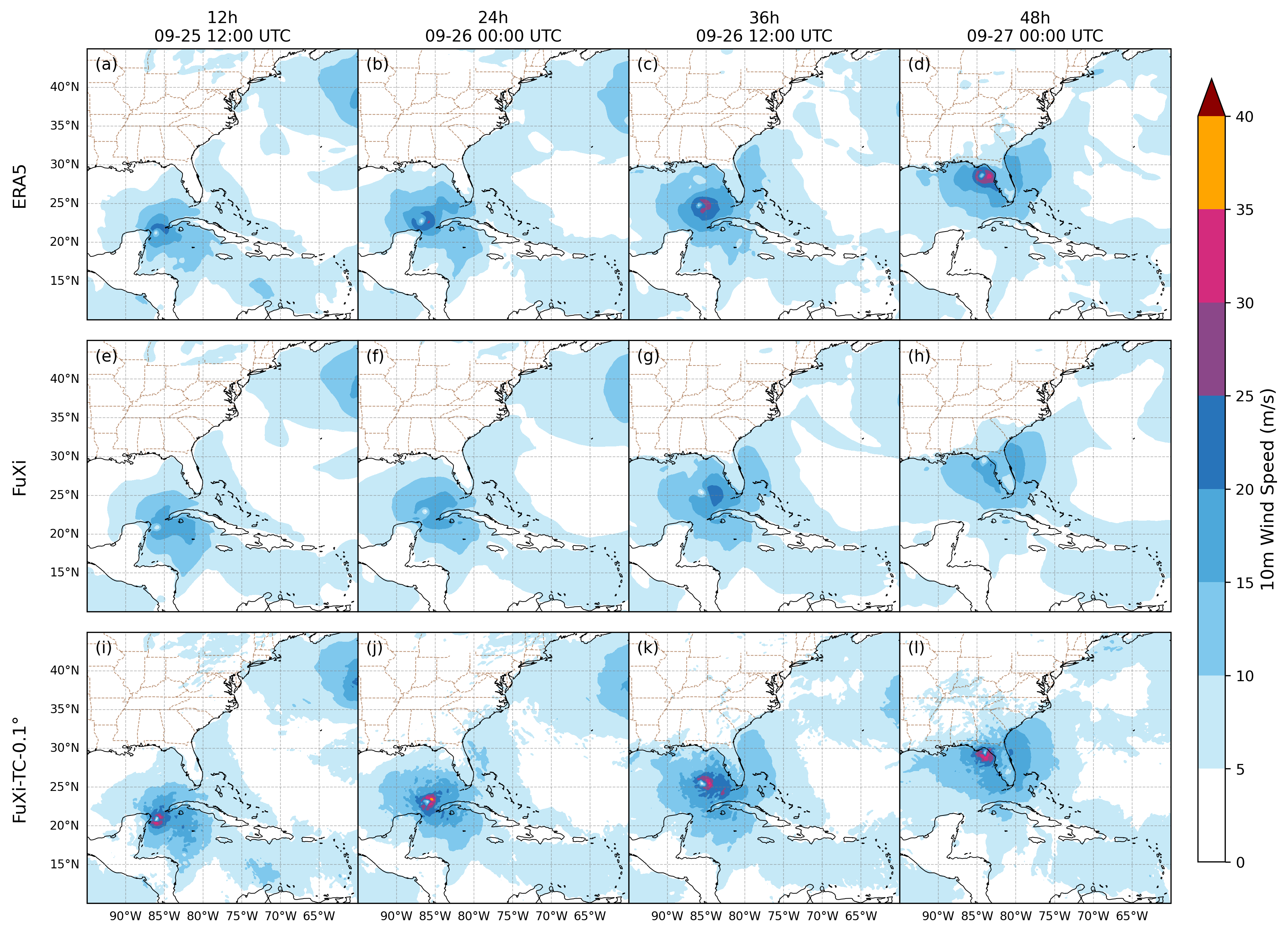}
    \caption{Comparison of WS10M for Hurricane Helene forecast. Comparison of snapshot examples of WS10M among ERA5 (first row), FuXi (second row) and FuXi-TC-\(0.1^\circ\) (third row) for 12 (first column), 24 (second column), 36 (third column) and 48 (fourth row) hours forecasts with the initial forecasting time at 00:00 UTC, 25 September, 2024.}
    \label{atlantic_helene}
\end{figure}

We present the forecast evolution characteristics of WS10M for Hurricanes Beryl (initial forecasting time at 12:00 UTC, 2 July 2024) and Helene (initial forecasting time at 00:00 UTC, 25 September 2024) in Figures~\ref{atlantic_beryl} and \ref{atlantic_helene}. The three rows of the figures sequentially compare the forecast snapshots of ERA5, FuXi, and FuXi-TC-\(0.1^\circ\), while the four columns correspond to the 12-, 24-, 36-, and 48-hour forecast results, respectively. It can be observed that FuXi-TC-\(0.1^\circ\) significantly increase the predicted values of WS10M while maintaining the consistency of the cyclone center structure with FuXi.

\FloatBarrier

\FloatBarrier

\section{Domain setup and TC track forecasts in the WNP and NA Basins}

\begin{figure}[t]
    \centering
    \includegraphics[width=\linewidth]{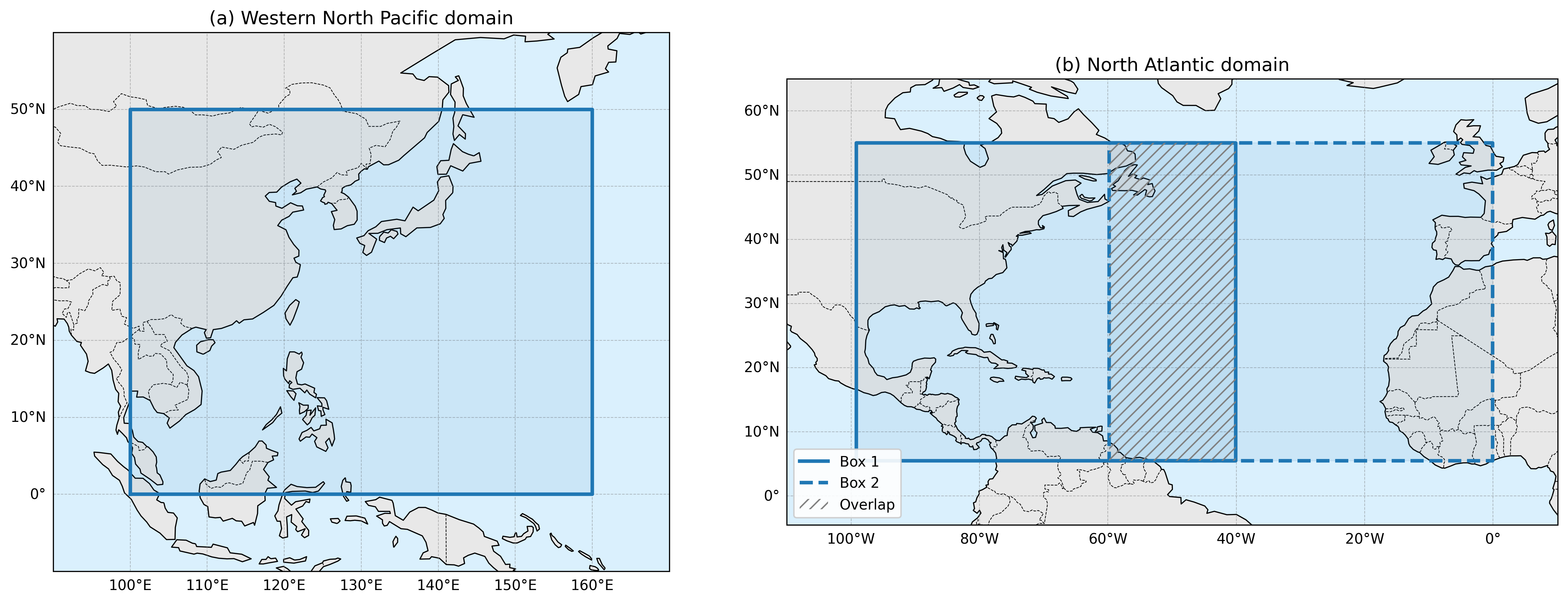}
    \caption{Domains used in this study. (a) The Western North Pacific (WNP) domain, (b) The North Atlantic (NA) domain. The NA domain is constructed by horizontally concatenating two overlapping sub-domains with identical size (Box 1, solid; Box 2, dashed), with the hatched area indicating the overlap region.}
    \label{domain_configuration}
\end{figure}

Supplementary Figure~\ref{domain_configuration} illustrates the domain configurations used for the WNP and NA basins. To accommodate the broader longitudinal span of the NA basin while maintaining consistent input dimensions, the NA domain is constructed by horizontally concatenating two sub-domains. Each of these NA sub-domains is identical in size to the WNP domain. 
Specifically, the western NA sub-domain spans \(99.2^\circ\)W--\(40.1^\circ\)W and \(5.5^\circ\)N--\(55^\circ\)N while the eastern NA sub-domain spans \(59.8^\circ\)W--\(0.1^\circ\)W and \(5.5^\circ\)N--\(55^\circ\)N.
In the overlapping region between the two NA sub-domains, the final meteorological fields are obtained by calculating the average of the overlapping grid points.

\FloatBarrier

\begin{figure}[t]
    \centering
    \includegraphics[width=\linewidth]{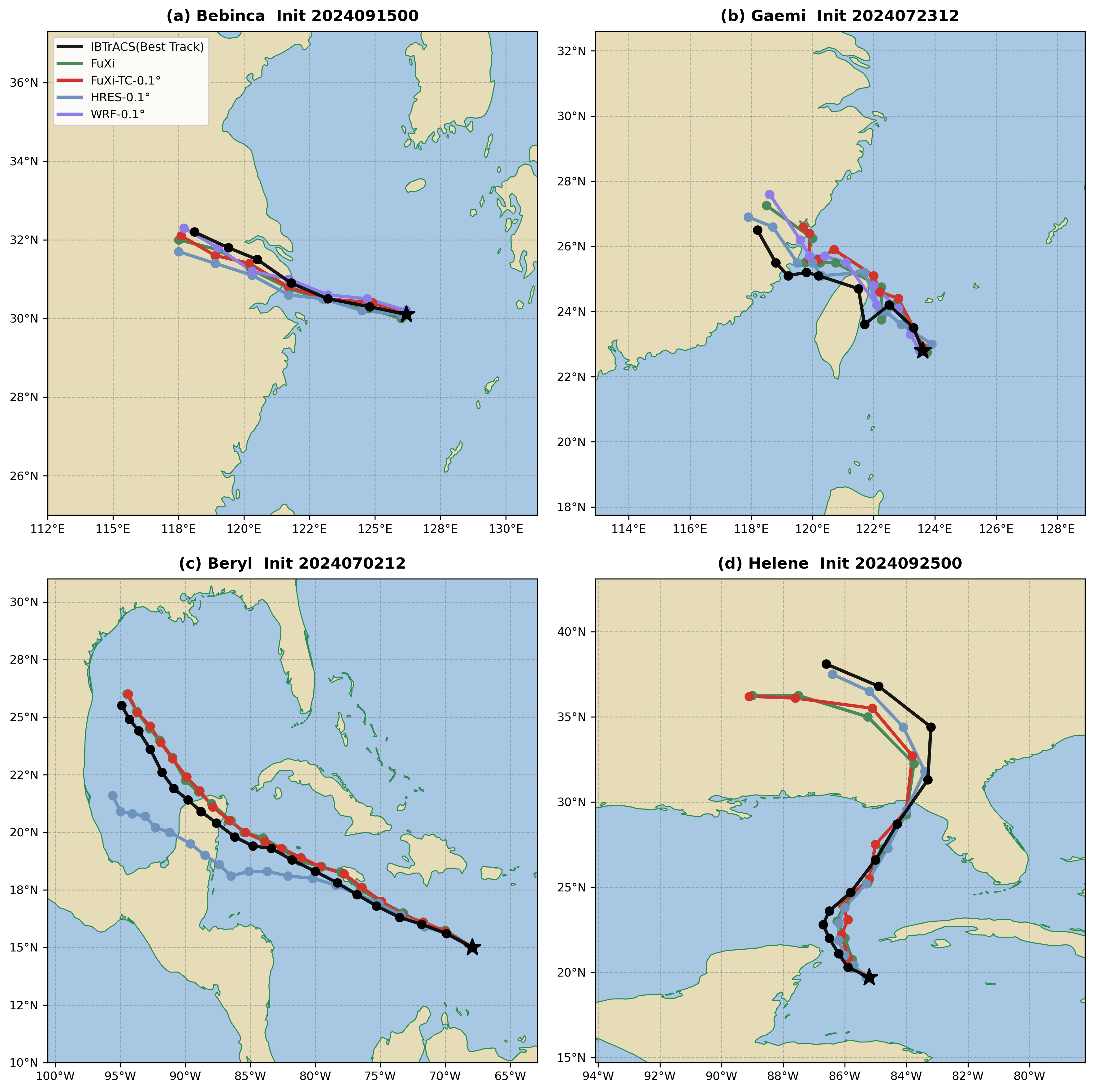}
    \caption{Track comparison of Typhoons (a) Bebinca, (b) Gaemi (c) Beryl and (d) Helene. Tracking results from FuXi (green line), FuXi-TC-\(0.1^\circ\) (red line), HRES-\(0.1^\circ\) (light blue line), and the observed Best Track (black line) are presented as lines with markers.}
    \label{track_comparison_4_cases}
\end{figure}

Within these defined domains, supplementary Figure~\ref{track_comparison_4_cases} presents the track forecast comparisons for four selected tropical cyclones cases. The first row displays Typhoons Bebinca and Gaemi over the WNP, while the second row shows Hurricanes Beryl and Helene over the NA. The results indicate that the track predictions of FuXi-TC-\(0.1^\circ\) (red line) are nearly identical to those of the baseline FuXi model (green line). For the late-stage forecasts of Hurricane Beryl, the deviation of HRES-\(0.1^\circ\) (light blue line) from the International Best Track (black line) gradually increases with the extension of the forecast lead time. Regarding Hurricanes Helene, the track forecasts generated by the FuXi-series models exhibit slightly larger deviations in the predicted landfall location.

\FloatBarrier

\bibliography{refs}

\noindent

\clearpage